\documentclass[a4paper,verbose,letterpaper,tmargin=2.5cm,bmargin=2.5cm,lmargin=2.5cm,rmargin=2.5cm]{article}

\usepackage[utf8]{inputenc}
\usepackage{url}
\usepackage{color}
\usepackage{lineno,setspace}
\usepackage{graphicx}
\usepackage{floatrow}
\usepackage{subfig}
\usepackage{authblk}
\usepackage{lipsum}
\usepackage{natbib}
\usepackage{amsmath,amsthm,amssymb,amsfonts,latexsym,multirow,verbatim}
\usepackage[normalem]{ulem}
\usepackage{lipsum}  
\usepackage{appendix}

\usepackage{tikz}
\usepackage{verbatim}
\usetikzlibrary{arrows,shapes}
\usetikzlibrary{decorations.pathmorphing}
\usetikzlibrary{decorations.pathreplacing}
\usetikzlibrary{decorations.shapes}
\usetikzlibrary{decorations.text}
\usetikzlibrary{decorations.markings}
\usetikzlibrary{decorations.fractals}
\usetikzlibrary{decorations.footprints}
\usepackage{sectsty}

\usepackage{geometry}

\usepackage{setspace}
\usepackage[english]{babel}
\usepackage{float}
\usepackage[T1]{fontenc}
\usepackage{amsmath}
\usepackage{amssymb}
\usepackage{amsfonts}
\usepackage{blkarray}
\usepackage{booktabs,caption}
\usepackage[flushleft]{threeparttable}
\usepackage{adjustbox}
\usepackage{makecell}
\usepackage{url}
\usepackage{comment} 
\usepackage{amsthm}
\theoremstyle{definition}

\usepackage{tikz}
\usetikzlibrary{shapes,arrows,snakes}
\usetikzlibrary{backgrounds,fit, positioning}
\usepackage{bbold}
\newcommand{\Cross}{\mathbin{\tikz [x=1.4ex,y=1.4ex,line width=.2ex] \draw (0,0) -- (1,1) (0,1) -- (1,0);}}%
\usepackage[thinlines]{easytable}

\theoremstyle{plain}

\newtheorem{prop}{Proposition}

\newtheorem{defn}{Definition}
\newtheorem{thm}{Theorem}

\usepackage{totcount}
\newtotcounter{citnum} 
\def\oldbibitem{} \let\oldbibitem=\bibitem
\newtotcounter{bibitems}
\regtotcounter{table}

\def\bibitem{\stepcounter{citnum}\oldbibitem}

\usepackage{amsmath}
\usepackage{graphicx}
\usepackage[colorinlistoftodos]{todonotes}
\usepackage[colorlinks=true, allcolors=blue]{hyperref}
\usepackage{amssymb}
\DeclareMathOperator{\EX}{\mathbb{E}}

\title{
Assessing mutualistic metacommunity capacity by integrating spatial and interaction networks
}
\author[1,*]{Marc Ohlmann}
\author[1,2]{François Munoz}
\author[3]{François Massol}
\author[1]{Wilfried Thuiller}

\affil[1]{Univ. Grenoble Alpes, CNRS, Univ. Savoie Mont–Blanc, LECA, Laboratoire d'Ecologie Alpine, F-38000 Grenoble, France}
\affil[2]{Univ. Grenoble Alpes, CNRS, Liphy, Laboratoire Interdisciplinaire de Physique, F-38000 Grenoble, France}
\affil[3]{Univ. Lille, CNRS, Inserm, CHU Lille, Institut Pasteur de Lille, U1019 - UMR 9017 - CIIL - Center for Infection and Immunity of Lille, F-59000 Lille, France}
\affil[*]{Corresponding author: marc.ohlmann@univ-grenoble-alpes.fr}
\date{}

\begin{document}


\maketitle

\section*{Abstract}
We develop a spatially realistic model of mutualistic metacommunities that exploits the joint structure of spatial and interaction networks. This model exhibits a sharp transition between a stable non-null equilibrium state and a global extinction state. This behaviour allows defining a threshold on colonisation/extinction parameters for the long-term metacommunity persistence. This threshold, the 'metacommunity capacity', extends the metapopulation capacity concept and can be calculated from the spatial and interaction networks without simulating the whole dynamics. In several applications we illustrate how the joint structure of the spatial and the interaction networks affects metacommunity capacity. It results that a weakly modular spatial network and a power-law degree distribution of the interaction network provide the most favourable configuration for the long-term persistence of a mutualistic metacommunity. Our model that encodes several explicit ecological assumptions should pave the way for a larger exploration of spatially realistic metacommunity models involving multiple interaction types.\newpage


\section{Introduction}
A fundamental goal of predictive ecology is to forecast the dynamics of interacting species in a given region (\citealt{thuiller2013road}, \citealt{mouquet2015predictive}). Reaching such a goal has direct implications for biodiversity management and conservation and to anticipate or mitigate the effects of habitat destruction and global change on biodiversity. 
\par
Metapopulation models have long been used to characterise the dynamics of populations that can colonise, persist or go extinct in a given landscape configuration (\citealt{hanski2003metapopulation}). This configuration is often summarised by a spatial network of suitable patches (\citealt{dale2010graphs,hagen2012biodiversity}) that best represents habitat patchiness in both natural and human-altered ecosystems (\citealt{haddad2015habitat}).
\citet{levins1969some} devised a seminal model of species occupancy \textit{i.e.}, the probability of presence of species populations across landscape. In this model, a mean-field, deterministic differential equation model represented the population dynamics in fully connected patches, so that equilibrium occupancy depended on both a colonisation and an extinction parameter. More than 30 years later, \citet{etienne2002non} proposed a stochastic analogue of Levins' model and studied the links between the properties of the two models.
Two sources of spatial heterogeneity can be embedded in metapopulation models: the heterogeneity on colonisation/extinction parameters among species (functional connectivity) and on the spatial network structure (structural connectivity) (\citealt{tischendorf2000usage}).
The impact of structural connectivity on stationary occupancy (\textit{e.g.}, \citealt{gilarranz2012spatial}) underlines the influence of fragmentation on metapopulation persistence (\citealt{fahrig2003effects}, \citealt{fletcher2018habitat}).
Subsequent deterministic, spatially realistic models acknowledged variation of connectivity among nodes, and allowed quantifying analytically the viability of a metapopulation that depends on the mere structural properties of the spatial network (\citealt{ovaskainen2001spatially}, \citealt{hanski2003metapopulation}). The viability is defined through the metapopulation capacity, \textit{i.e.}, a threshold on colonisation and extinction parameters above which the metapopulation can survive. This threshold is thus of prime importance in biological conservation (\citealt{groffman2006ecological}).
\par
However, species populations are likely to interact with many other species within habitat patches. These interactions should also affect the spatial coexistence of multiple metapopulations and their respective capacities (\citealt{thuiller2013road}).
Metacommunity models are designed to assess the joint dynamics of multiple species populations in an habitat network (\citealt{leibold2004metacommunity}).
While the structure of interaction networks is known to strongly influence biodiversity dynamics (\citealt{sole2007}), most existing deterministic metacommunity models generally focused on global competition and competition-colonisation trade-off in fully connected patches (\citealt{tilman1997habitat}, \citealt{calcagno2006coexistence}), or sometimes in evenly connected patches (\textit{e.g.}, lattice \citealt{amarasekare2004mechanisms}, \citealt{mouquet2011extinction}).
Models focusing on other interaction types (e.g. mutualistic and trophic) were developed  for species-poor communities (\textit{i.e.} two species, \citealt{nee1997two} or for few species \citealt{gravel2020toward}), preventing the study of complex networks and further generalisations.
\par
Yet, stochastic models of interactions where species are either present or absent can encode mechanisms through specific rules, like having at least one prey to survive in the Trophic Theory of Island Biogeography (\citealt{Gravel2011}, \citealt{Massol2017}), or through increasing probability of presence depending on prey availability in a model originally designed for network inference (\citealt{auclair2017labeled}). The latter model belongs to graphical models, a class of statistical models that represents conditional dependencies between species distributions using graphs. Using network-based metrics, these models can encode several mechanisms in terms of conditional probabilities of presence (\cite{staniczenko2017linking}). Nevertheless, these approaches still ignore the spatial structure of the environment.
\par
So far, theoretical studies on the dynamics of metacommunities within a spatially explicit environment and with biotic interactions have never considered how the dynamics jointly depend on graph properties of both interaction and spatial networks (e.g., \citealt{amarasekare2004mechanisms}), trophic interactions (\citealt{pillai2010patch}, \citealt{brechtel2018master}, \citealt{gross2020modern} but see \citealt{wang2021metapopulation}) or mutualistic interactions on a lattice (\citealt{filotas2010effect}, \citealt{sardanyes2019habitat}). These models often elude the question of existence of a non-null equilibrium, and the metacommunity persistence is often assessed through tedious dynamic simulations or using strong approximations (\citealt{wang2021metapopulation}). If this approach provides points in the parameter space where the metacommunity persists, it neither maps regions of this space leading to persistence, nor it demonstrates the existence of critical thresholds acting on metacommunity persistence as in metapopulation theory.
\par
Interestingly, thresholds between local community persistence and extinction have already been identified in the case of positive interactions (\citealt{callaway1997positive}, \citealt{kefi2016can}). For instance, mutualistic interactions play a major role in natural systems by conditioning coexistence (\citealt{valdovinos2019mutualistic}). \citet{thebault2010stability} showed that mutualistic networks generally have a nested architecture favouring persistence, and empirical surveys evidenced  a truncated power-law of degree distribution  (\citealt{bascompte2006structure}, \citealt{vazquez2009evaluating}, \citealt{bascompte2009mutualistic}). However, no network-based model of spatially realistic, mutualistic metacommunities has been proposed so far. Such model should allow to test the joint impact of the structure of the spatial and interaction networks on the viability of a metacommunity and, potentially, allow to exhibit thresholds acting at the mutualistic metacommunity level. It should also reconcile the ongoing debate on the impact of the structure of the spatial network on metapopulations (\citealt{fletcher2018habitat}).
\par
In this paper, we explicitly model mutualistic interactions in an heterogeneous space using dynamic Bayesian networks (\citealt{auclair2017labeled}). We derive the mean-field model and show a threshold in metacommunity persistence, defining an abrupt transition between stable coexistence and global metacommunity extinction. Our approach extends the computation of metapopulation capacity \textit{sensu} \citeauthor{ovaskainen2001spatially} to the case of mutualistic metacommunities. Using numerical methods, we show how metacommunity capacity relies on the structure of both mutualistic and spatial networks. Importantly, specific submodels can be derived to encode key ecological assumptions on extinction and colonisation. For these different ecological assumptions, we represent how spatial proximity of sites and mutualistic interactions modulate colonisation and/or extinction probability, and we derive metacommunity capacities. We finally explore the relationship between the degrees of the nodes of both spatial and interaction networks and species' occupancy at equilibrium. This allows extracting ecological relevant quantities on species among the sites (\textit{e.g.} mean occupancy) or in sites between species (\textit{e.g.} species diversity, interaction network diversity).
We thus show that metacommunity viability can be understood in the light of the joint structure of spatial and interaction networks.
\\

 \section{Stochastic models of metacommunity dynamics using dynamic Bayesian networks}
 We first present a formalism that unifies spatially realistic metapopulation models and mainland-island models of biotic interactions in discrete time using Dynamic Bayesian Networks (DBNs). DBNs describe dependencies between random variables at different time steps through a bipartite directed graph, and represent stochastic models in which parameters are networks (\citealt{lahdesmaki2008learning}, \citealt{koller2009probabilistic}).
 Given a set of $n$ random variables $(X_1$,...,$X_n)$ (we note $\text{I}=\{1,...,n\}$),
 \begin{defn}
  Two random variables $X_i$ and $X_j$ are independent conditionally given $\mathbf{X_{I \smallsetminus \{i,j\}}}$  iff:
$$
     \mathbb{P}(X_i,X_j|\mathbf{X_{I \smallsetminus \{i,j\}}}) = \mathbb{P}(X_i|\mathbf{X_{I \smallsetminus \{i,j\}}})\mathbb{P}(X_j|\mathbf{X_{I \smallsetminus \{i,j\}}})
$$
 \end{defn}
\noindent
Bayesian networks aim to map conditional independence statements using a Directed Acyclic Graph $G$ (DAG). For a given node $u$, we note $Pa_u(G)$ the set of nodes that are parents of $u$.
\begin{equation}
    Pa_u(G)=\{v \in V, (v,u)\in E\}
\end{equation}
\noindent    
The joint probability $\mathbb{P}(\mathbf{X})$ factorises over $G$ as :
\begin{equation}
    \mathbb{P}(X_1,...,X_n) = \prod_i \mathbb{P}(X_i|\mathbf{X}_{Pa_{i}(G)})
\end{equation}
The factorisation gives the independence conditional statement according to the structure of the DAG.

\noindent
A particular case of Bayesian network consists in Dynamic Bayesian Networks (DBNs). Indexing our previous $n$ random variables by time $t$, a DBN describes the homogeneous dependencies between $\{X_1^t,...,X_n^t\}$ and $\{X_1^{t+1},...,X_n^{t+1}\}$ using a directed bipartite network $G_{bip}$ (we note $\mathbf{A_{bip}}$ its adjacency matrix). Importantly, as the structure of $G_{bip}$ does not depend of $t$, it can be built using an aggregated network $G$ (we note $\mathbf{A}$ its adjacency matrix) and a graph $P_2$ (we note $\mathbf{A_2}$ its adjacency matrix) whose set of nodes is $\{t,t+1\}$ and set of edges is $\{(t,t+1)\}$. We have
\begin{equation}
        \mathbf{A_{bip}} = \mathbf{A_2} \otimes (\mathbf{A} + \mathbf{I_n})
\end{equation}
\noindent
where $\mathbf{I_n}$ denotes the identity matrices of dimension $n$. We set $\mathbf{\Tilde{A}} = \mathbf{A} + \mathbf{I_n}$ and denotes $\Tilde{G}$ the associated graph.
The joint probability factorizes over $G_{bip}$:
\begin{equation}
        \mathbb{P}(X_1^{t+1},...,X_n^{t+1}|X_1^t,...,X_n^t) = \prod_i \mathbb{P}(X_i^{t+1}|\mathbf{X}_{Pa_{i}(\Tilde{G})})
\end{equation}

\noindent
 The network represents the causal influences between species distributions between two time steps.
 Once the structure of causal influences is fixed, several distributions can be associated to a given network structure through different parameterisations.
 These parameterisations represent interaction mechanisms that describe the effect of neighbour species or sites on the probability of presence of a given species at time $t+1$.
 \\
The heterogeneous space is represented by a spatial network $G_s=(V_s,E_s)$. We assume that this network is undirected and connected, \textit{i.e.}, considering two nodes $u$ and $v$ of $G_s$, there is always a path from $u$ to $v$. Biotic interactions in the metacommunity are represented by an interaction network $G_b=(V_b,E_b)$, which we also assume undirected and connected. We note $n=|V_s|$ and $m=|V_b|$ (see Table \ref{tab_not} for notations).\\


 \begin{table}[!htbp]
\caption{Notations}
\begin{adjustbox}{width=\textwidth}
\renewcommand{\arraystretch}{1.2}
\begin{tabular}{|c|l|}
\hline
\textbf{Object}                           & \textbf{Name} \\ \hline
$G_s$     & Spatial network ($n$ nodes)              \\ \hline
$G_b$                        & Interaction network ($m$ nodes) \\ \hline  
$G_s^0$     & Spatial network where edges have been deleted ($n$ nodes)              \\ \hline
$G_b^0$                        & Interaction network where edges have been deleted ($m$ nodes) \\ \hline
$G_{s,b} = G_s \square{} G_b$                      & Cartesian product of the spatial and biotic interaction networks ($n*m$ nodes)               \\ \hline  
$\mathbf{A_s}$                              & Adjacency matrix of the spatial network           \\ \hline  
$\mathbf{A_b}$                              & Adjacency matrix of the biotic interaction network            \\ \hline
$\mathbf{A_{s,b}} = \mathbf{A_s} \otimes \mathbf{I_m} + \mathbf{I_n} \otimes \mathbf{A_b}$                               &  Adjacency matrix of the Cartesian product network)
\\ \hline
$G_c$                               &  Colonisation network ($n*m$ nodes)
\\ \hline
$G_e$                               &  Extinction network ($n*m$ nodes)
\\ \hline
$\mathbf{A_c} $                               &  Adjacency matrix of the colonisation network
\\ \hline
$\mathbf{A_e}$                               &  Adjacency matrix of the extinction network \\ \hline
$\lambda_M$                               &  Metacommunity persistence capacity
\\ \hline
$\lambda_I$                               &  Metacommunity invasion capacity  \\ \hline  
$\Lambda_s$                              & Dominant eigenvalue of the adjacency matrix of the spatial network
\\ \hline  
$\Lambda_b$                              & Dominant eigenvalue of the adjacency matrix of the biotic interaction network
\\ \hline
$\Lambda_{s,b} = \Lambda_s + \Lambda_b$                               &  Dominant 
eigenvalue of the adjacency matrix of the Cartesian product network
 \\ \hline
\end{tabular}
\renewcommand{\arraystretch}{1.2}
\end{adjustbox}
\label{tab_not}
\end{table}


\subsection{Spatially realistic metapopulation model}
\label{section:spom}

Let $X_i^t$ be a random variable associated to the presence of a population in a site $i$ (\textit{i.e.} the node $v_i$ of $G_s$) at time $t$ ($1 \leq i \leq n$, $t \in \mathbb{N^*}$). We depict the dependency structure between the $X_i^t$ using a DBN built from $G_s$ (Fig. \ref{fig:models}a). 
Defining the neighbours of $v_i$ in $G_s$ as $N_s(i)$, the parents of $X_i^{t+1}$ in the DBN are $\{X_i^t,\mathbf{X_{N_s(i)}^t}\}$. This means that the presence of a population at time $t+1$ is causally influenced by the presence of a population at time $t$ in site $i$ and in sites adjacent to $i$.
In this first model, no other variables or species influence the presence of a population in site $i$ at time $t+1$.
Through conditional probabilities, the parameterisation encodes the way the presence or absence of a population in adjacent sites modulates the probability of presence of a population in the focal site. Here, we chose the same parameterisation as in \citealt{gilarranz2012spatial}.
\begin{equation}
  \label{metapop}
     \mathbb{P}(X_i^{t+1}=1| X_i^t,\mathbf{X}_{N_s(i)}^t)=(1-(1-c)^{\sum_{k \in N_s(i)}{X_k^t}})(1-X_i^t) + (1-e)X_i^t  
\end{equation}
 where $c$ and $e$ are the respective colonisation $(0<c<1)$ and extinction $(0<e<1)$ parameters. In Eq. \ref{metapop}, the probability of presence grows with the number of occupied adjacent sites. Specifically, the probability that node $i$ includes a population at time $t+1$ is $1-e$ if it had one at time $t$, while the probability that node $i$ is colonised between time $t$ and time $t+1$ is equal to $1$ minus the probability that all occupied neighbouring sites do not colonise node $i$, which happens with probability $1-c$ independently for each of these nodes. \\
 The spatially realiscic metapopulation model is a homogeneous Markov chain on $\chi = \{0,1\}^n$.
 A state of the metapopulation is a binary vector of length $n$ indicating whether each site is occupied or not.
 The dimension of the transition matrix is $2^n*2^n$ and the probability of transition between a state $s_k=(x_1,...,x_n)$ and $s_l=(\Tilde{x_1},...,\Tilde{x_n})$ is 
 \begin{equation}
     P_{k,l}=\mathbb{P}(X_1^{t+1}=x_1,...,X_n^{t+1}=x_n|X_1^{t}=\Tilde{x_1},...,X_n^{t}=\Tilde{x_n})
 \end{equation}
 \begin{equation}
     P_{k,l}=\prod_i \mathbb{P}(X_i^{t+1}=x_i|X_i^{t}=\Tilde{x_i},\mathbf{X_{N_s(i)}^t}=(\Tilde{x_{N_s(i)}}))
 \end{equation}
 $\mathbf{0}$ is an absorbing state of the model. However, the model will reach a quasi-stationary distribution (see \citealt{darroch1965quasi}) before extinction which gives a distribution of all possible states of the metapopulation among sites. Getting extinction time and quasi-stationary distribution require to compute eigenvectors and eigenvalues of $\mathbf{P}$ that are intractable in the general case since $\mathbf{P}$ is high-dimensional.
 
 \subsection{A mainland-island model with biotic interactions}
 \label{section:mainland_island}
  Dynamic Bayesian networks can also be used to build mainland-island stochastic models of species interactions.
 Let $X_j^t$ be the random variable associated to the presence of population of species $j$ on the island. A DBN representing the dependency structure is  built from $G_b$ (Fig. \ref{fig:models}a). Here, the DBN represents the network of species interactions as links affecting the probability that a species present on the island goes extinct, or that an absent species is able to colonise the island.
 Defining as $N_{G_b}(j)$ the neighbours of $v_j$ in $G_b$ , the parents of $X_j^{(t+1)}$ in the DBN are $\{X_j^t,\mathbf{X_{N_{G_b}(j)}^t}\}$, meaning that the presence of species $v_j$ and species that interact with $v_j$ at time $t$ on the island, causally influences the presence of species $v_j$ at time $t+1$. Importantly, there is no other variables influencing the presence of a species $v_j$ at time $t+1$. We chose a parameterisation similar to \citealt{auclair2017labeled}:
 \begin{equation}
    \mathbb{P}(X_{j}^{t+1}|X_{j}^t,\mathbf{X}_{N_{G_b}(j)}^t)=c(1-X_{j}^t) + (1- e(1 - \frac{\sum_{k \in N_{G_b}(j)} X^t_{k}}{1+deg_{G_b}(j)}))X_{j}^t
     \label{mainlandisland}
 \end{equation}
where $deg_{G_b}(j)$ is the degree of $j$ in $G_b$. The probability of extinction (defined by Eq. \ref{mainlandisland}) belongs to $]0,1[$ (Appendix). 
Although the dependency between species occurrences can encode any kind of interactions, we here focus on the mutualistic case by imposing an extinction function. In this case, the probability of extinction of a given species decreases with the number of species present that interact with the focal species. \\
The mainland-island model of species interaction is a homogeneous Markov chain on $\chi = \{0,1\}^m$ with no absorbing state. 
 A state of the mainland-island model of species interaction is a binary vector of length $m$, representing the composition of the community. The dimension of the transition matrix is $2^m*2^m$ and the probability of transition between a state $s_k=(x_1,...,x_m)$ and $s_l=(\Tilde{x_1},...,\Tilde{x_m})$ is 
 \begin{equation}
     P_{k,l}=\prod_j \mathbb{P}(X_j^{t+1}=x_j|X_j^{t}=\Tilde{x_j},X_{N_b(j)}^t=(\Tilde{x_{N_b(j)}}))
 \end{equation}
 The chain converges towards a unique stationary distribution, a distribution of probability over all possible species communities. However, as in the metapopulation case, computing the stationary distribution is intractable in the general case since $\mathbf{P}$ is high-dimensional.

\noindent
To summarise, in the metapopulation model, the spatial network acts on the probability of colonisation, whereas in the interaction model, the biotic network acts on the probability of extinction. \\
 
 \subsection{Spatially realistic models of mutualistic metacommunities}
 \label{section:sto_metacom}
 Integrating the models from Section \ref{section:spom} and \ref{section:mainland_island}, we built a spatially explicit metacommunity model using $G_s$ and $G_b$. To do so, we used the Cartesian product of graphs that builds a network from $G_b$ and $G_s$ (\citealt{imrich2000product}).
 
 \begin{figure}
\centering
\includegraphics[scale = 0.6]{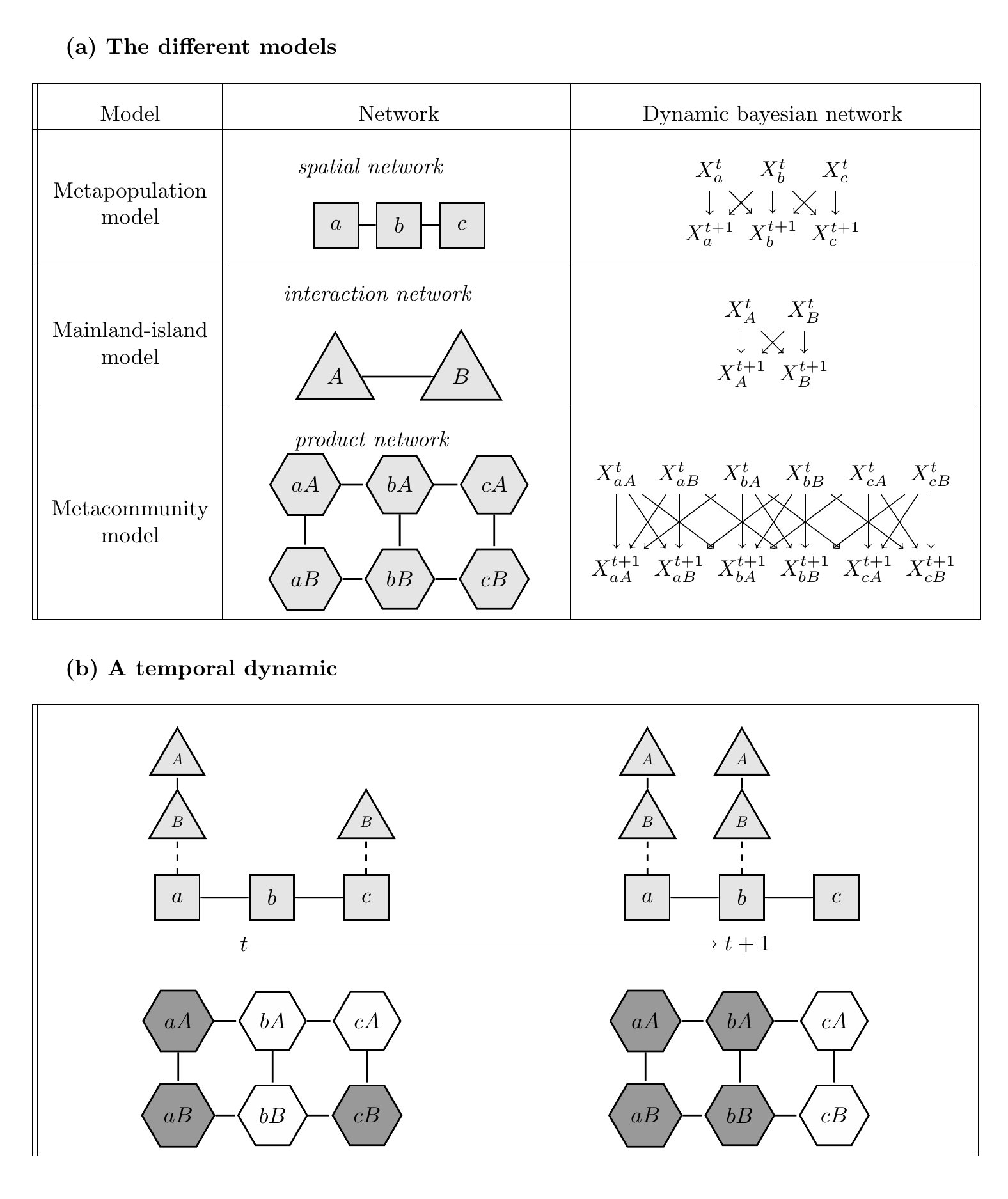}
\caption{
(a) Metapopulation model, mainland-island interaction model and metacommunity model. The second column represents the network associated to each model (spatial, interaction and product network). The third column represents the dynamic Bayesian network associated to each model that represents the causal influences of variables (presence of populations of a given species, species on the island, species in sites) at $t$ on variables at $t+1$ \\
(b) Simulating a dynamic in the combined effect model between two time steps. The nodes of the product network are either empty or occupied (grey: occupied, white: empty). For the sake of simplicity, the model here is turned deterministic ($c=1$, $e=1$). To colonise a new node of the product network, species $A$ and $B$ must be both present in the same site and can colonise adjacent site only. The population of species $B$ originally present in site $c$ goes extinct since it does not co-occur with $A$ at $t$ whereas species $A$ and $B$ that co-occur in site $a$ colonise the site $b$.}
\label{fig:models}
\end{figure}

 \begin{defn}
 The cartesian product of $G_s$ and $G_b$, $G_{s,b}=G_s \square G_b$ is the graph in which the set of nodes is $V_d \Cross V_b$. A node of this graph is identified by a pair of nodes of $G_s$ and $G_b$. Moreover, there is an edge between $(u_s,u_b)$ and $(v_s,v_b)$ if ($u_s=v_s$ and $(u_b,v_b) \in E_s$) or ($u_b=v_b$ and $(u_s,v_s) \in E_b$).
\end{defn}
\noindent The adjacency matrix, $A_{s,b}$, of $G_{s,b}$ is 
\begin{equation}
\mathbf{A_{s,b}} = \mathbf{A_s} \otimes \mathbf{I_m}+ \mathbf{I_n} \otimes \mathbf{A_b}
\end{equation}
where $\mathbf{I_m }$ and $\mathbf{I_n}$ denotes the identity matrices of dimension $m$ and $n$ and $\otimes$ denotes the Kronecker product of two matrices.

 Let $X_{ij}^t$ be the random variable associated to the presence of a population of species $j$ in site $i$ at time $t$.
 The dependency structure between the $X_{ij}^t$ is depicted using a DBN that is built from $G$ (Fig. \ref{fig:models}a). Defining as $N(i,j)$ the neighbours of $(v_i,v_j)$ in $G_{s,b}$, the parents of $X_{ij}^{t+1}$ in the DBN are $\{X_{ij}^t,\mathbf{X_{N(i,j)}^t}\}$. This means that the presence of a population of species $j$ in site $i$ at time $t+1$ is causally influenced by the presence of population of the same species in adjacent sites at time $t$ and by the presence of populations of species that interact with $j$ in the same site. In the mainland-island model of interaction, we assumed that colonisation probability is constant and that interactions act on the extinction probability contrary to the metapopulation model. \\
 At that stage, this is crucial to define several submodels that formalise key ecological assumptions in the product graph, using either the spatial network or the colonisation network to modulate colonisation and/or extinction probability. \\
Let $G_s^0$ be the network that has the same set of nodes as $G_s$ but an empty set of edges, and let  $G_b^0$ be the network that has the same set of nodes as $G_b$ but an empty set of edges. We introduce then the colonisation network $G_c$ ($\mathbf{A_c}$ is its adjacency matrix) and the extinction network $G_e$ ($\mathbf{A_e}$ is its adjacency matrix). These networks modulate the colonisation and extinction probability in the different submodels. We build four submodels from a given product graph (Table \ref{tab_models}) :
 \begin{itemize}
     \item a Levins type submodel, where both the spatial and biotic interaction networks modulate the colonisation probability ($G_c=G_s \square G_b$), while the extinction probability is constant ($G_e=G_s^0 \square G_b^0$)
     \item a separated-effect submodel, where the spatial network modulates the colonisation probability ($G_c=G_s \square G_b^0$) and the biotic interaction network modulates the extinction probability ($G_e=G_s^0 \square G_b$)
     \item a combined effect submodel, where both the spatial and the biotic interaction networks modulate the colonisation probability ($G_c=G_s \square G_b$), and the biotic interaction network modulates the extinction probability ($G_e=G_s^0 \square G_b$)
     \item a rescue effect submodel, where the spatial network modulates the colonisation probability ($G_c=G_s \square G_b^0$) and both the spatial and the biotic interaction networks modulate the extinction probability ($G_e=G_s \square G_b$).
 \end{itemize}
For the four submodels, the conditional probabilities of colonisation and non-extinction are expressed as:
 \begin{equation}
     \mathbb{P}(X_{ij}^{t+1}=1|X_{ij}^t=0, \sum_{(k,l) \in N_{G_c}(i,j)} X^t_{kl})=\epsilon + (1-\epsilon)*(1-(1-c)^{\sum_{(k,l) \in N_{G_c}(i,j)} X^t_{kl}})
\end{equation}
\begin{equation}
    \mathbb{P}(X_{ij}^{t+1}=1|X^t_{kl}=1, \sum_{(k,l) \in N_{G_e}(i,j)} X^t_{kl})=1-e(1 - \frac{\sum_{(k,l) \in N_{G_e}(i,j)} X^t_{kl}}{1+deg_{G_e}((i,j))})
\end{equation}
where $\epsilon \in ]0;1[$ is a constant that guarantees the convergence of the model. This constant allows colonisation from an external source, analogous to nodal self-infection in the epidemiology literature (\citealt{van2012epidemics}). The proposed metacommunity model is analogous to the open Levins model, that better fits with data than the classic Levins model (\citealt{laroche2018abundance}).
$deg_{G_e}((i,j))$ is the degree of $(v_i,v_j)$ in $G_e$, and $N_{G_c}(i,j)$ (resp. $N_{G_e}(i,j)$) denotes the neighbours of $(v_i,v_j)$ in $G_c$ (resp. $G_e$). Fig. \ref{fig:models}b shows a simplistic dynamics in the combined effect model.
\begin{prop}
The stochastic spatially realistic metacommunity model converges towards a unique stationnary distribution
\end{prop}
 In the stochastic spatially realistic models of mutualistic metacommunities, the transition matrix of the chain is of dimension $2^{mn}*2^{mn}$, encoding the probability of transition between a state $\mathbf{s_k}=(x_{11},...,x_{mn})$ of the metacommunity and a state $\mathbf{s_l}=(\Tilde{x_{11}},...,\Tilde{x_{mn}})$, where $x_{ij} \in \{0,1\}$ describes the presence of a population of species $i$ in site $j$.
We note $\mathbf{P}$ the transition matrix, the probability of transition between $\mathbf{s_k}$ and $\mathbf{s_l}$ is :
\begin{equation}
         P_{k,l}=\prod_{i,j} \mathbb{P}(X_{ij}^{t+1}=x_{ij}|X_{11}^{t}=\Tilde{x_{11}},...,X_{mn}^{t}=\Tilde{x_{mn}})
\end{equation}
Moreover, we have:
\begin{equation}
        \mathbb{P}(X_{i,j}^{t+1}=1|X_{i,j}^t=0, \sum_{(k,l) \in N_{G_c}(i,j)} X^t_{k,l})=\epsilon + (1-\epsilon)*(1-(1-c)^{\sum_{(k,l) \in N_{G_c}(i,j)} X^t_{k,l}})
\end{equation}
with $\epsilon \in ]0;1[$ and $c \in ]0;1[$. We have:
\begin{equation}
 \mathbb{P}(\epsilon<\epsilon + (1-\epsilon)*(1-(1-c)^{\sum_{(k,l) \in N_{G_c}(i,j)} X^t_{k,l}})<1)=1
\end{equation}
Moreover:
\begin{equation}
    \mathbb{P}(X_{i,j}^{t+1}=1|X^t_{i,j}=1, \sum_{(k,l) \in N_{G_e}(i,j)} X^t_{k,l})=1-e(1 - \frac{\sum_{(k,l) \in N_{G_e}(i,j)} X^t_{k,l}}{1+deg_{G_e}((i,j))})
\end{equation}
where $e \in ]0;1[$.
We have then:
\begin{equation}
\frac{e}{1+deg_{G_e}((i,j))}<e(1 - \frac{\sum_{(k,l) \in N_{G_e}(i,j)} X^t_{k,l}}{1+deg_{G_e}((i,j))})<e
\end{equation}
The probability of extinction is in $]0;1[$.\\
Consequently : 
\begin{equation}
    \forall i \in \{1,...,n\},\forall j \in \{1,...,m\}, \  \mathbb{P}(X_{i,j}^{t+1}=x_{i,j}|X_{1,1}^{t}=\Tilde{x_{1,1}},...,X_{m,n}^{t}=\Tilde{x_{m,n}}) > 0
\end{equation}
It follows that $\mathbf{P}$ is irreducible and $(\mathbf{X^t})^t$ converges towards a unique stationary distribution. Importantly, in the stationnary distribution, each species in each sites has a non-nul probability of presence. Computing the stationary distribution is also intractable in the general case (since transition matrix is of dimension $2^{nm}$), but, it is however possible to simulate the dynamics of the metacommunity as \cite{gilarranz2012spatial} did for metapopulation model.

 \section{The $nm$-intertwined model}
 Since studying the stochastic model of Section \ref{section:sto_metacom} is intractable in the general case, we propose to study deterministic models that approximate the stochastic models, referred to as the intertwined model in the epidemiology literature (\citealt{van2011n}). We extended the spatially realistic Levins model to the product of spatial and interaction network (\citealt{ovaskainen2001spatially}). The approximation is derived from \cite{bianconi2018multilayer} and \cite{van2011n}. The aim is to study the dynamics of occupancy of each species $j$ in each site $i$, \textit{i.e.} $p_{ij}(t)=\EX(X_{ij}^t)=\mathbb{P}(X_{ij}^t=1)$. For all $i$ and $j$:
 \begin{equation}
    p_{ij}(t+1)=\EX((1-X_{ij}^t)(\epsilon + (1-\epsilon)(1-(1-c)^{\sum_{(k,l) \in N_{G_c}(i,j)} X^t_{kl}})))+\EX((1-e(1-e)^{\sum_{(k,l) \in N_{G_e}(i,j)} X^t_{kl}})X_{ij}^t)
    \label{raweq}
 \end{equation}
Eq. \ref{raweq} leads to a hierarchy of equations that cannot be solved (\textit{i.e.} we need to consider $\EX(X^t_{1,1},...,X^t_{m,n})$ to find a solution to the system). A useful approximation consists in the mean field approximation that assumes, for any sequence of indices $n(1),n'(1);...,n(r),n'(r')$ :
\begin{equation}
   \EX(X^t_{n(1),n'(1)},...,X^t_{n(r),n'(r')}) \simeq \EX(X^t_{n(1),n'(1)})...\EX(X^t_{n(r),n'(r')})
\end{equation}
 After some algebra, introducing a new single index $v$ for the nodes of the product network and assuming that $c=o(1)$,$e=o(1)$ and $\epsilon=o(c)$ (see Appendix), it follows :
\begin{equation}
    p_v(t+1)-p_v(t)=C_v(\mathbf{p(t)})(1-p_v(t)))-E_v(\mathbf{p(t)})(p_v(t))
    \label{eq:deter_model}
\end{equation}
where $C_v(\mathbf{p(t)})=c\sum_{u}[A_c]_{v,u}p_{u}(t)$ and $E_v(\mathbf{p(t)})=e(1-\sum_{u}[A_e]_{v,u}p_{u}(t)/M_u)$ with $M_u = 1 +deg_{G_e}(u)$ \\
This rewrites:
\begin{equation}
\label{eq:master}
\mathbf{p(t+1)}-\mathbf{p(t)} = c(\mathbf{A_c}\mathbf{p(t)})\odot (\mathbf{1} - \mathbf{p(t)}) -e(\mathbf{1} - (\mathbf{D_e} + \mathbf{I}_{nm})^{-1}\mathbf{A_e}\mathbf{p(t)}) \odot \mathbf{p(t)} 
\end{equation}

\noindent
where $\odot$ denotes the element-wise product, $\mathbf{D_e}$ denotes the in-degree matrix of $G_e$ and $\mathbf{I}_{nm}$ denotes the identity matrix of dimension $nm$. \\
Eq. \ref{eq:deter_model} is analogous to master equation of \cite{ovaskainen2001spatially}. Now, to assess the viability of a given mutualistic metacommunity, we need to determine the equilibrium states and evaluate their local stability. 

 \subsection{A recall on metapopulation capacity} 
 \label{section:metapop_cap}

In metapopulation models, equilibrium state is either stable coexistence (all sites have non-null occupancy) or global extinction (all patches have null occupancy). Metapopulation capacities have thus been derived to assess both the persistence and the stability of metapopulations  at equilibrium (\citealt{hanski2000metapopulation}, \citealt{ovaskainen2001spatially}). The metapopulation persistence capacity $\lambda_M$ is a break-point between coexistence and global extinction, a threshold (scalar quantity) computable from the spatial network.
More formally, in the metapopulation case, $G_b$ is made of a single node, ($m=1$) and we assume that $G_s$ is undirected and connected. We have:
\begin{equation}
    \forall t \in \mathbb{N}^*, \ \mathbf{p(t)}\in \overline{\Omega} =  \{x \in \mathbb{R}^n, \forall i, \  0\leq x_i \leq 1 \}
\end{equation}
\noindent
with the following assumptions on the colonisation functions (per site $i$), $C_i(.)$, and extinction functions, $E_i(.)$:
\begin{itemize}
    \item there is no external source of migrants
    \begin{equation}
    C_i(\mathbf{0})=0
    \end{equation}
    \item the occupied sites make a positive contribution to the colonisation function of an empty site
    \begin{equation}
    \forall \mathbf{p} \in \Omega = \{x \in \mathbb{R}^n, \forall i, \  0< x_i < 1 \}, C_i(\mathbf{p})>0
    \end{equation}
    \begin{equation}
    \begin{cases}
    \frac{\partial C_i}{\partial p_j}(\mathbf{p})  \geq 0  & \text{for} \,  i \neq j  \\
    \frac{\partial C_i}{\partial p_i}(\mathbf{p}) =0 &  
    \end{cases}
    \end{equation}
    \item there is no mainland population, extinction rates are positive and, eventually, reduced by the presence of local populations
    
    \begin{equation}
    \forall p \in \overline{\Omega}, E_i(\mathbf{p})>0 
    \end{equation}
    
    \begin{equation}
    \begin{cases}
    \frac{\partial E_i}{\partial p_j} \leq 0 & \text{for}   \, i \neq j  \\ \frac{\partial E_i}{\partial p_i}=0 &
     \end{cases}
    \end{equation}
    
    \item Colonisation and extinction functions are smooth functions
    \begin{equation}
    C_i \in \mathcal{C}^1(\overline{\Omega})
    \end{equation}
    \begin{equation}
    E_i \in \mathcal{C}^1(\overline{\Omega})
    \end{equation}
\end{itemize}
Let:
\begin{equation}
    g_i(\mathbf{p}) = \frac{eC_i(\mathbf{p})}{cE_i(\mathbf{p})}
\end{equation}
The model is also assumed to be irreducible. Let $\mathbf{J}$ be the matrix of dimension $n\times n$ so that:
\\
\begin{equation}
J_{ij} = \begin{cases}
1 &\text{ if }  \frac{\partial g_i}{\partial p_j}(\mathbf{p}) \geq 0, \mathbf{p} \in \Omega  \\
0 &\text{ otherwise }
\end{cases}
\end{equation}
We say that the model is irreducible if $\mathbf{J}$ is irreducible, i.e. the graph that has $\mathbf{J}$ as adjacency matrix is strongly connected.\\
In the case of the spatially realistic Levins model :
\begin{itemize}
\item $C_i(\mathbf{p}) = c(\mathbf{A_{s}}\mathbf{p})_i$
\item $E(\mathbf{p}) = e$
\end{itemize}
where $\mathbf{A_{s}}$ is the adjacency matrix of the spatial network. Then,
\begin{equation}
    g_i(\mathbf{p}) = (\mathbf{A_{s}}\mathbf{p})_i
\end{equation}
and the model is irreducible since $\mathbf{A_{s}}$ is irreducible.\\
The metapopulation inviasion capacity, $\lambda_I$, is defined as the dominant eigenvalue of the jacobian matrix of $g$ evaluated in $\mathbf{p}=0$. It measures the stability of the equilibrium $\mathbf{p}=0$ that is the ability of a single population to invade the spatial network.\\
\begin{defn}
The metapopulation persistence capacity, $\lambda_M$, is defined as:
$$
    \lambda_M = \sup_{\mathbf{p}\in \Omega} h(\mathbf{p})
$$
where 
$$
    h(\mathbf{p})= \min_i h_i(\mathbf{p})
$$
and
$$
    h_i(\mathbf{p})= g_i(\mathbf{p})\frac{1-p_i}{p_i}
$$
\end{defn}

We now present a weak version of the main theorem of \cite{ovaskainen2001spatially}.
\begin{thm}
(Ovaskainen \& Hanski) The deterministic metapopulation
model has a nontrivial equilibrium state
if and only if the threshold condition $\lambda_M > \frac{e}{c}$ (if all the components of $\mathbf{g}$ are concave) or
$\lambda_M \geq \frac{e}{c}$ (otherwise) is satisfied
\label{thm:ova}
\end{thm}
$\lambda_M$ is a threshold on the colonisation/extinction parameters that allows the metapopulation to persist. Importantly, if the metapopulation persists, the equilibrium point is interior (it belongs to $\Omega$), meaning that all occupancies are strictly positive. \\
Moreover, if all the components of $\mathbf{g}$ are concave (it is the case for the spatially realistic Levins model), we have:
\begin{equation}
    \lambda_M=\lambda_I=\Lambda_s
\end{equation}
where $\Lambda_s$ is the dominant eigenvalue of $\mathbf{A_{s}}$.
If one component (or more) of $\mathbf{g}$ is not concave, then $\lambda_I < \lambda_M$

\subsection{Extension to mutualistic metacommunity capacity}
\subsubsection{The mutualistic metacommunity concept}
We extend metapopulation capacities from Section \ref{section:metapop_cap} to mutualistic metacommunity capacities in the dynamical system defined by Eq. \ref{eq:master}, using the product of the spatial network and the biotic interaction network and specific assumptions on colonisation and extinction functions. The proposed mutualistic metacommunity model presents a sharp transition between coexistence (all species have non-null occupancy in all sites) and global extinction (all species have null occupancy in all sites).\\
In this case, $\Omega = \{x \in \mathbb{R}^{n*m}, \forall v \in \{1,...,n*m \}\  0< x_v < 1 \}$\\
We have:
    \begin{equation}
        C_v(\mathbf{p(t)})=c\sum_{u}[A_{c}]_{v,u}p_{u}(t)
    \end{equation}
\noindent
and   
    \begin{equation}
        E_v(\mathbf{p(t)})=e(1-\sum_{u}[A_{e}]_{v,u}p_{u}(t)/M_u))
    \end{equation}

\noindent
In order to apply theorem \ref{thm:ova} to the product network, we first verify assumptions on colonisation and extinction functions (notice that index $v$ represents a combination of a site and a species index).
\begin{itemize}
    \item there is no external source of migrants
    \begin{equation}
        C_v(\mathbf{0})=0
    \end{equation}
    Notice that this assumption is only verified at order $1$
    \item species occupying sites make a positive contribution to the colonisation function of an empty site
    \begin{equation}
    \forall \mathbf{p} \in \Omega, C_v(\mathbf{p})>0
    \end{equation}
    \begin{equation}
    \begin{cases}
    \frac{\partial C_v}{\partial p_u}(\mathbf{p})  \geq 0  & \text{for} \,  u \neq v  \\
    \frac{\partial C_v}{\partial p_v}(\mathbf{p}) =0 &  
    \end{cases}
    \end{equation}
    \item there is no mainland population, extinction rates are positive and reduced by the presence of others species
    \begin{equation}
    \forall p \in \overline{\Omega}, E_v(\mathbf{p})>0 
    \end{equation}
    \begin{equation}
    \begin{cases}
    \frac{\partial E_i}{\partial p_j} \leq 0 & \text{for}   \, i \neq j  \\ \frac{\partial E_i}{\partial p_i}=0 &
     \end{cases}
    \end{equation}
    Notice that, due to this assumption, we stick to the modelling of mutualistic metacommunity.
    
    \item Colonisation and extinction are smooth functions
    \begin{equation}
    C_v \in \mathcal{C}^1(\overline{\Omega})
    \end{equation}
    \begin{equation}
    E_v \in \mathcal{C}^1(\overline{\Omega})
    \end{equation}
\end{itemize}
Additionally:
\begin{prop}
\label{prop:irreducible}
The four proposed metacommunity submodels are irreducible
\end{prop}

See proof in Appendix. \\
We then define metacommunity invasion capacity as the dominant eigenvalue of the jacobian matrix of $g$ evaluated in $\mathbf{p}=0$.\\

\begin{defn}
The metacommunity persistence capacity, $\lambda_M$, is defined as:
$$
    \lambda_M = \sup_{\mathbf{p}\in \Omega} h(\mathbf{p})
$$
where 
$$
    h(\mathbf{p})= \min_v h_v(\mathbf{p})
$$
and
$$
    h_v(\mathbf{p})= g_v(p)\frac{1-p_v}{p_v}
$$
\end{defn}

By applying theorem \ref{thm:ova}, a non-trivial equilibrium that the dynamical system has a non trivial equilibrium if and only if $\lambda_M > \frac{e}{c}$. $\lambda_M$ is then a threshold on the colonisation/extinction parameters that allows the mutualistic metacommunity to persist. Importantly, a non-trivial equilibrium point is interior (it belongs to $\Omega$), so each species in each site has a positive abundance at equilibirum. \\
\begin{prop}
For the Levins type model, $\lambda_M = \lambda_I = \Lambda_s+\Lambda_b$
\end{prop}

The Levins type model is actually the spatially realistic model with $G_c = G_s \square G_b$ as spatial network. The dominant eigenvalue, $\Lambda_c$, of $A_c$ is $\Lambda_s+\Lambda_b$. Consequently, for the Levins type submodel:
\begin{equation}
    \lambda_M = \lambda_I = \Lambda_s+\Lambda_b
\end{equation}
$\lambda_I$ for the four different models are provided in Appendix.
\subsubsection{Computation of metacommunity capacity}
Metacommunity capacities can be computed for each of the four different submodels. The Levins type model (Table \ref{tab_models}) is actually a spatially realistic Levins model where the product network is analogue to the spatial network in the classic spatially realistic metapopulation model. In this case, $\lambda_M=\lambda_I=\Lambda_b+\Lambda_c$. Notice that in this case, both the mutualistic and the spatial networks play interchangeable roles. For the combined effect model, the separated effect model and the rescue effect model, we computed the metacommunity capacity $\lambda_M$ using Appendix D of \citealt{ovaskainen2001spatially} and simulating annealing. We propose an implementation in R and Python. The code to compute the metacommunity capacity in the different models is available at: \url{https://gitlab.com/marcohlmann/metacommunity_theory}.
Note that in general, only the metapopulation or the metacommunity persistence capacity is really the focus. For the sake of simplicity, we will thus use metacommunity capacity as metacommunity persistence capacity in the rest of the text (unless specified otherwise) 


\begin{table}[!htbp]
\caption{Top: Map of the different models, submodels and their parameters. Bottom: The four submodels associated to mutualistic metacommunity models, their assumptions, colonisation/extinction networks and metacommunity capacities}
\begin{adjustbox}{width={\textwidth}}
\begin{tabular}{l}
\tikzstyle{decision} = [rectangle, draw, fill=gray!05, 
    text width=7em, text badly centered, node distance=2cm, inner sep=2pt]

\tikzstyle{spatial} = [scale=0.3,regular polygon,regular polygon sides = 4
, draw, fill=gray!05, 
    text width=3.5em, text badly centered, node distance=2cm, inner sep=0pt,fill = gray!30]   

    \tikzstyle{interaction} = [scale=0.2,regular polygon,regular polygon sides = 3
, draw, fill=gray!05, 
    text width=3.5em, text badly centered, node distance=2cm, inner sep=0pt,fill = gray!30]   
 
    \tikzstyle{product} = [scale=0.3,regular polygon,regular polygon sides = 6
, draw, fill=gray!05, 
    text width=3.5em, text badly centered, node distance=2cm, inner sep=0pt,fill = gray!30]    
    
\tikzstyle{block} = [rectangle, text=white, draw, fill=black, 
    text width=8em, text centered, rounded corners, minimum height=4em,font=\bfseries]
 
    \tikzstyle{sub} = [rectangle, text=white, draw, fill=black, 
    text width=8em, text centered, rounded corners, minimum height=2em,font=\bfseries]

\tikzstyle{line} = [draw, -latex']
\tikzstyle{edgebis}=[snake=expanding waves,segment length=1mm,segment angle=20,draw]

\tikzstyle{cloud} = [draw, ellipse,fill=gray!20, node distance=3cm,
    minimum height=2em,text width=4em]

\tikzstyle{colo} = [draw, rectangle,fill=gray!20,double = black, node distance=3cm,
    minimum height=1.5em,text width=5.2em,
rounded corners,inner sep=2pt,text centered]

\tikzstyle{exti} = [draw, rectangle,fill=gray!20,double = black, node distance=3cm,
    minimum height=1.5em,text width=5.2em,
rounded corners,text centered]
    
\begin{tikzpicture}[node distance = 2cm, auto]
    \node [block]  at (0,0) (DBN) {DBN formalism};
    \node [spatial,label={below:\small spatial network}] at (2,2) (spatial) {};
    \node [interaction,label={\small interaction network}] at (2,-2) (interaction) {};
    \node [product,label={\small product network}] at (6.5,0) (product) {};
    \node [block] at (3.5,0) (metacom) {metacommunity models};
    \node [block] at (-2,2) (metapop) {metapopulation models};
    \node [block] at (-2,-2) (mainland-island) {mainland-island models};
    \node [decision] at (10,3) (lev) {Levins type submodel};
    \node [spatial] at (10,2.3) (spatial_lev) {};
    \node [interaction] at (10,1.65) (inter_lev) {};
    
    \node [decision] at (10,1) (comb) {Combined effect model};
    \node [spatial] at (10,0.3) (spatial_comb) {};
    \node [interaction] at (10,-0.35) (inter_comb) {};
    
    \node [decision] at (10,-1) (sep) {Separated effect submodel};
    \node [spatial] at (10,-1.7) (spatial_sep) {};
    \node [interaction] at (10,-2.35) (inter_sep) {};
    
    \node [decision] at (10,-3) (resc) {Rescue effect submodel};
    \node [spatial] at (10,-3.7) (spatial_resc) {};
    \node [interaction] at (10,-4.35) (inter_resc) {};
    
    \node [colo] at (2,3.5) (colo) {colonisation};
    \node [exti] at (2,-3.5) (exti) {extinction};
    
    \node [colo] at (14,1.5) (colo_bis) {colonisation};
    \node [exti] at (14,-1.5) (exti_bis) {extinction};
    \node at (16,-1.5) (detour) {};
    \node [sub] at (10,4) (submodels) {submodels}; 

    \path [edgebis] (DBN) -- (mainland-island);
    \path [edgebis] (DBN) -- (metacom);
    \path [edgebis] (DBN) -- (metapop);
    \path [line,dashed,thick] (metapop) --  (spatial) ;
    \path [line,dashed,thick] (mainland-island) --  (interaction) ;
    \path [line,dashed,thick] (metacom) --  (product) ;
    \path [line,dashed,thick] (spatial) edge[bend left] node [left] {} (product);
     \path [line,dashed,thick] (interaction) edge[bend right] node [left] {} (product);
     \path [line,dotted,thick] (product) edge[bend left] node [left] {} (lev);
    \path [line,dotted,thick] (product) edge node [left] {} (comb);
    \path [line,dotted,thick] (product) edge node [left] {} (sep);
    \path [line,dotted,thick] (product) edge[bend right] node [left] {} (resc);
    \draw[line]  (spatial) -- node[sloped,pos=0.5,text width=2em] {acts on} (colo); 
    \draw[line]  (interaction) -- node[sloped,pos=0.5,text width=2em] {acts on} (exti);  
 
    \draw[line]  (spatial_lev) edge[bend left= 10] (colo_bis); 
    \draw[line]  (inter_lev) edge[bend left=10] (colo_bis);
    
    \draw[line]  (spatial_comb) edge[bend right= 10] (colo_bis); 
    \draw[line]  (inter_comb) edge[bend right= 10] (colo_bis);
    \draw[line]  (inter_comb) edge[bend left= 20] (exti_bis);
    
     \draw[line]  (spatial_sep) edge[bend right= 35] (colo_bis); 
    \draw[line]  (inter_sep) edge[bend right= 10] (exti_bis);
    
    \draw[line]  (spatial_resc) to[out=0,in=-90] (detour) to[out=90,in=0] (colo_bis); 
    \draw[line]  (inter_resc) edge[bend right= 30] (exti_bis);
    \draw[line]  (spatial_resc) edge[bend right= 30] (exti_bis);

\end{tikzpicture}
\end{tabular}
\end{adjustbox}
\begin{adjustbox}{width={\textwidth}}
\begin{tabular}{|l|l|l|l|}
\cline{1-4}
 \makecell{\textbf{Name of the submodel}}  & \makecell{\textbf{Mechanisms and assumptions}} & \makecell{\textbf{Colonisation and extinction networks}} & \makecell{\textbf{Metacommunities capacities}}   \\ \cline{1-4}
\large Levins type & \makecell{The spatial and the biotic network modulate the colonisation probability.\\ The extinction probability is constant.\\ The different sites and species acts independently  \\  on the probability of presence of a species.}  & \makecell{\Large $G_c = G_s \square G_b$ \\\Large $G_e =G_s^0 \square G_b^0$} & \makecell{\Large $\lambda_I=\lambda_M= \Lambda_s + \Lambda_b$}  \\ \cline{1-4}
\large Combined effect & \makecell{The spatial and the biotic network modulate the colonisation probability.\\ The biotic interaction network modulates the extinction probability.\\ The different sites and species acts independently \\ on the probability of presence of a species.} & \makecell{\Large $G_c = G_s \square G_b$ \\\Large $G_e =G_s \square G_b^0$} & \makecell{\Large$\lambda_I = \Lambda_s + \Lambda_b$ \\\Large $\lambda_M$: to compute numerically}   \\ \cline{1-4}
\large Separated effect & \makecell{The spatial network modulates the colonisation probability.\\  The biotic network modulates the extinction probability.\\ The different sites and species acts independently  \\  on the probability of presence of a species.} &  \makecell{\Large $G_c = G_s \square G_b^0$ \\\Large $G_e =G_s^0 \square G_b$} & \makecell{\Large $\lambda_I=\Lambda_s$ \\ \Large $\lambda_M$: to compute numerically}  \\ \cline{1-4}
\large Rescue effect & \makecell{The spatial network modulates the colonisation probability.\\  The spatial and the biotic network modulates the extinction probability.\\ The different sites and species acts independently  \\ on the probability of presence of a species.} &  \makecell{\Large $G_c = G_s \square G_b^0$ \\ \Large $G_e =G_s \square G_b$} & \makecell{\Large$\lambda_I=\Lambda_s$ \\ \Large$\lambda_M$: to compute numerically}   \\ \cline{1-4}
\end{tabular}
\end{adjustbox}
\label{tab_models}
\end{table}

 \section{Applications}
 \subsection{Illustration}
\begin{figure}
\centering
\includegraphics[width=\textwidth]{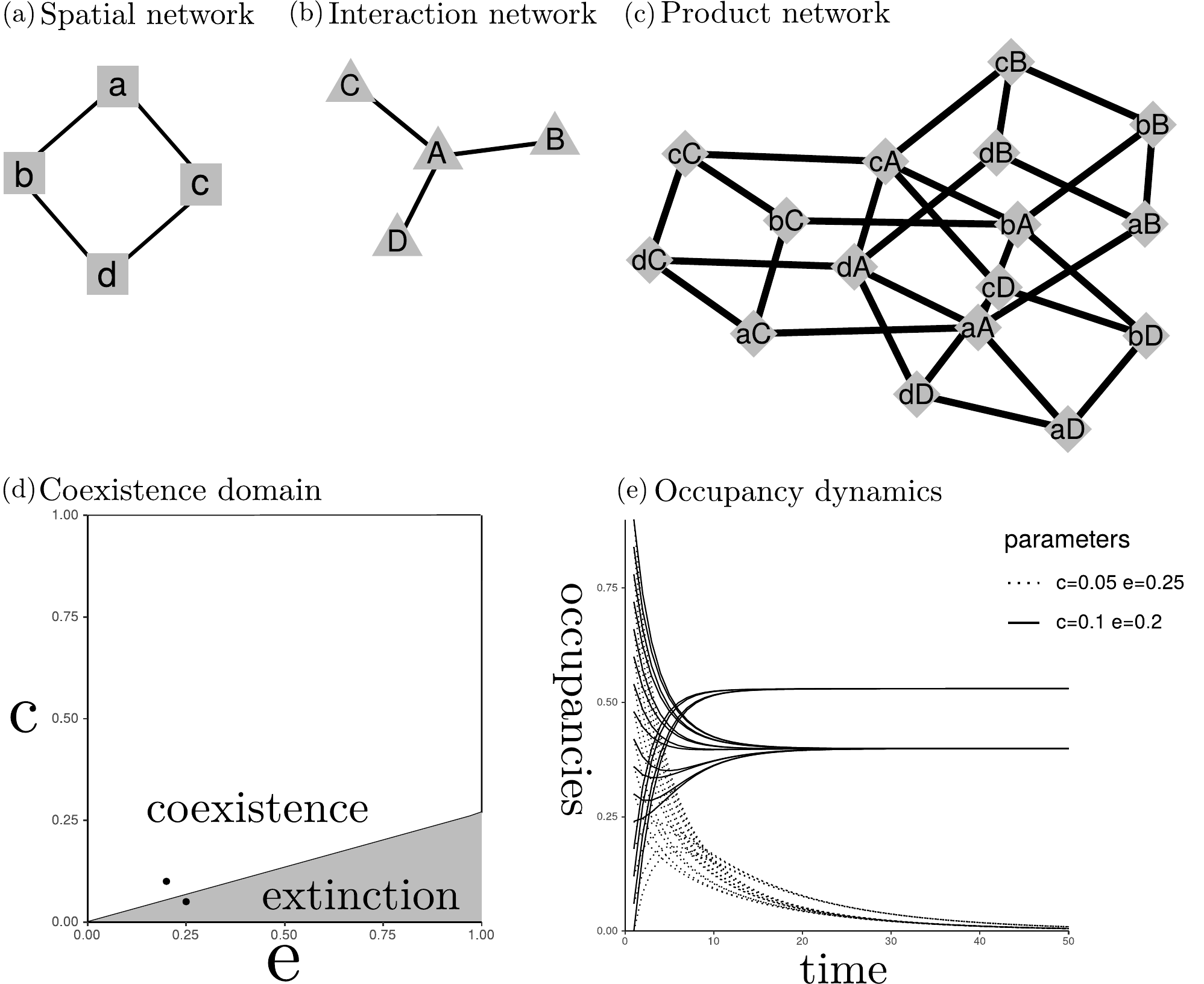}
\caption{Toy model built from (a) a circular spatial network, (b) a star-shaped interaction network giving (c) the product network. The product network defines (d) the persistence and extinction domain. (e) Two trajectories sampled in and outside the persistence domain leading to persistence or extinction of the metacommunity}
\label{fig:toy_model}
\end{figure}

To illustrate the metacommunity capacity concept, we built a toy model (Fig. \ref{fig:toy_model}). We used a circular spatial network with $4$ nodes (Fig. \ref{fig:toy_model}a) and a star shaped interaction network made of $4$ nodes (Fig. \ref{fig:toy_model}b), which could represent a plant species and its mutualistic mycorrhizal fungi species. The Cartesian product is built from the spatial and the interaction networks (Fig. \ref{fig:toy_model}c). For the illustration, we derived the Levins type submodel dynamics. In this case, both metacommunity invasion capacity $\lambda_I$ and persistence capacity $\lambda_M$ are equal to the dominant eigenvalue of the product of the networks ($3.73$). $\lambda_M$ defines the feasibility domain that is the portion of space where all species have a non-null abundance (see \cite{song2018guideline}) (Fig. \ref{fig:toy_model}d). We showed two possible outcomes of species occupancy dynamics (Fig. \ref{fig:toy_model}). One had a combination of colonisation and extinction values allowing metacommunity persistence, while the other had values outside the feasibility domain and yielded metacommunity extinction.
Occupancies of persisting species converge toward two different values due to symmetries in the product network. Despite its simplicity, this toy model shows that we can predict the outcome of mutualistic metacommunity dynamics for any location of the parameter space, depending on the metacommunity capacity.
 \subsection{Structures of spatial and mutualistic interaction network jointly shape the metacommunity capacity}
 
 We applied our model to investigate how the structure of the spatial and interaction networks shape the metacommunity capacity of a bipartite mutualistic system. 
To simulate landscape fragmentation, we sampled two types of spatial networks while keeping constant the expected number of edges. We generated random spatial networks with $10$ nodes in either Erd\H{o}s-Renyi graphs (all edges are independent and identically distributed, with connectance $C=0.25$) or modular graphs using a block model ($C=0.25$, more details in Appendix).
 We only kept connected spatial networks and used $15$ replicates for each type of spatial network.
 Concerning the mutualistic network, we sampled two types of bipartite networks while keeping constant the number of edges. We generated random interaction networks with $14$ nodes and $16$ edges in either Erd\H{o}s-Renyi graphs or networks with degree distribution shaped as a power-law of scaling parameter equals to $2$. We used the function \textit{sample\_fitness\_pl} implemented in the R package \textit{igraph} (\citealt{igraph}).
 We only kept connected interaction networks and used $15$ replicates per type of interaction network.
 We then computed the colonisation and extinction networks for each combination of spatial and interaction networks, so generating $4*4*15=900$ different networks in total. This number of replicates was large enough to generate robust results (see Appendix).
 We first computed the metacommunity capacities for each combination of spatial and interaction networks to assess the viability range of the metapopulations. Then, since a metacommunity will persist given the parameters, we studied how species occupancy at equilibrium and aggregated quantities build from these occupancies (mean occupancy, species diversity) depend on node characteristics of both networks.
 \subsubsection{Computing metacommunity capacities}
\begin{figure}[h!]
\centering
\includegraphics[width=\textwidth]{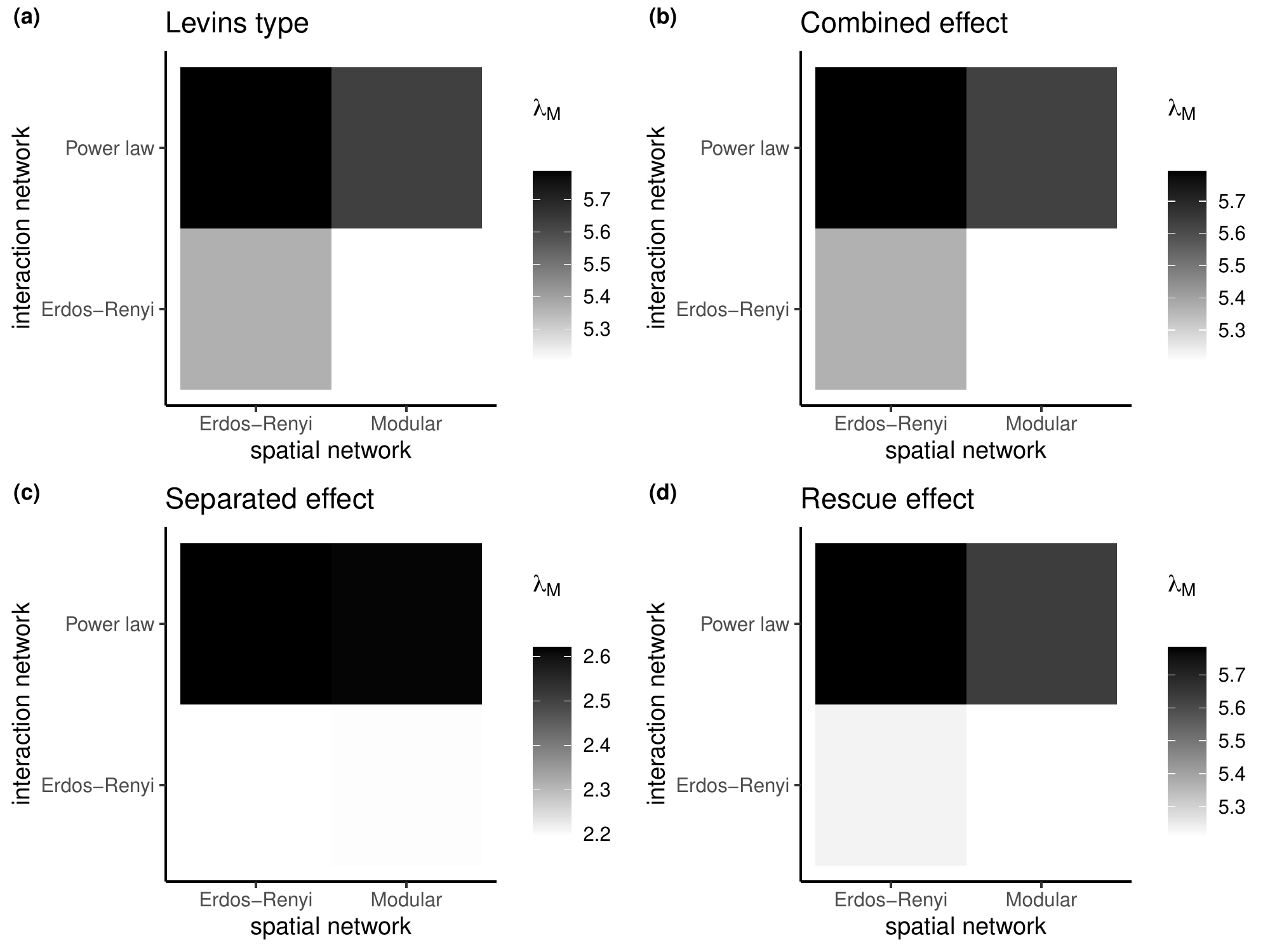}
\caption{Assessing metacommunity persistence capacity in function of the spatial and interaction networks. (a) Metacommunity persistence capacity for the Levins type submodel in function of the structure of the spatial networks (Erd\H{o}s-Renyi/Modular) and the structure of the interaction networks (Erd\H{o}s-Renyi/Power law) (b) Metacommunity persistence capacity for the combined effect submodel. (c) Metacommunity persistence capacity for the separated effect submodel. (d) Metacommunity persistence capacity for the rescue effect submodel.}
\label{fig:lambda_M_four_models}
\end{figure}

We computed the metacommunity capacity $\lambda_M$ for the four different submodels and the four combinations of networks structure (Fig. \ref{fig:lambda_M_four_models}). Despite known concerns on the ability to fit power-laws on small networks (\citealt{clauset2009power}, \citealt{stumpf2012critical}), we were able to statistically distinguish estimation of metacommunity capacity for almost all sampled combinations of structures (Appendix).
For the Levins type submodel, the metacommunity capacity decreased when the spatial network was modular and when the degree distribution was not a power-law. In this case, the modularity of the spatial network had a stronger impact on the metacommunity capacity than the structure of the mutualistic interaction network. For the combined effect submodel, we observed a similar trend than with the Levins type submodel. For the separated-effect submodel, metacommunity capacities were significantly lower than both Levins type and combined effect submodels. Moreover, the structure of the spatial network had little impact on the metacommunity capacity contrary to the structure of the interaction network. For the rescue effect submodel, the structure of the spatial network did not impact the metacommunity capacity when the structure of the interaction network was Erd\H{o}s-Renyi but, when the interaction network had a power-law degree distribution, modular spatial network decreased the metacommunity capacity. Importantly, for the separated and the rescue effect submodels, whatever the structure of the spatial network (modular or Erd\H{o}s-Renyi), mean $\lambda_M$ was not statistically different for a power-law or a Erd\H{o}s-Renyi mutualistic interaction network.

 \subsubsection{A focus on species occupancy at equilibrium for a given network combination}
\par
Here, we focused on a given network combination and chose extinction and colonisation values ensuring metacommunity persistence. We simulated metacommunity dynamics and studied how the occupancy at equilibrium of each node of the product network depends on its degree for the Levins type submodel (see Appendix for the other three submodels).
Additionally, we studied the mean occupancy of species across sites, plus species and link diversity in each site. We used a spatial network with modularity of $0.36$ and a mutualistic network with a degree distribution sampled in a power-law with parameter 2. We set the colonisation parameter to $c=0.1$ and the extinction parameter to $e=0.1$. 
\par
\begin{figure}
\centering
\includegraphics[width=\textwidth]{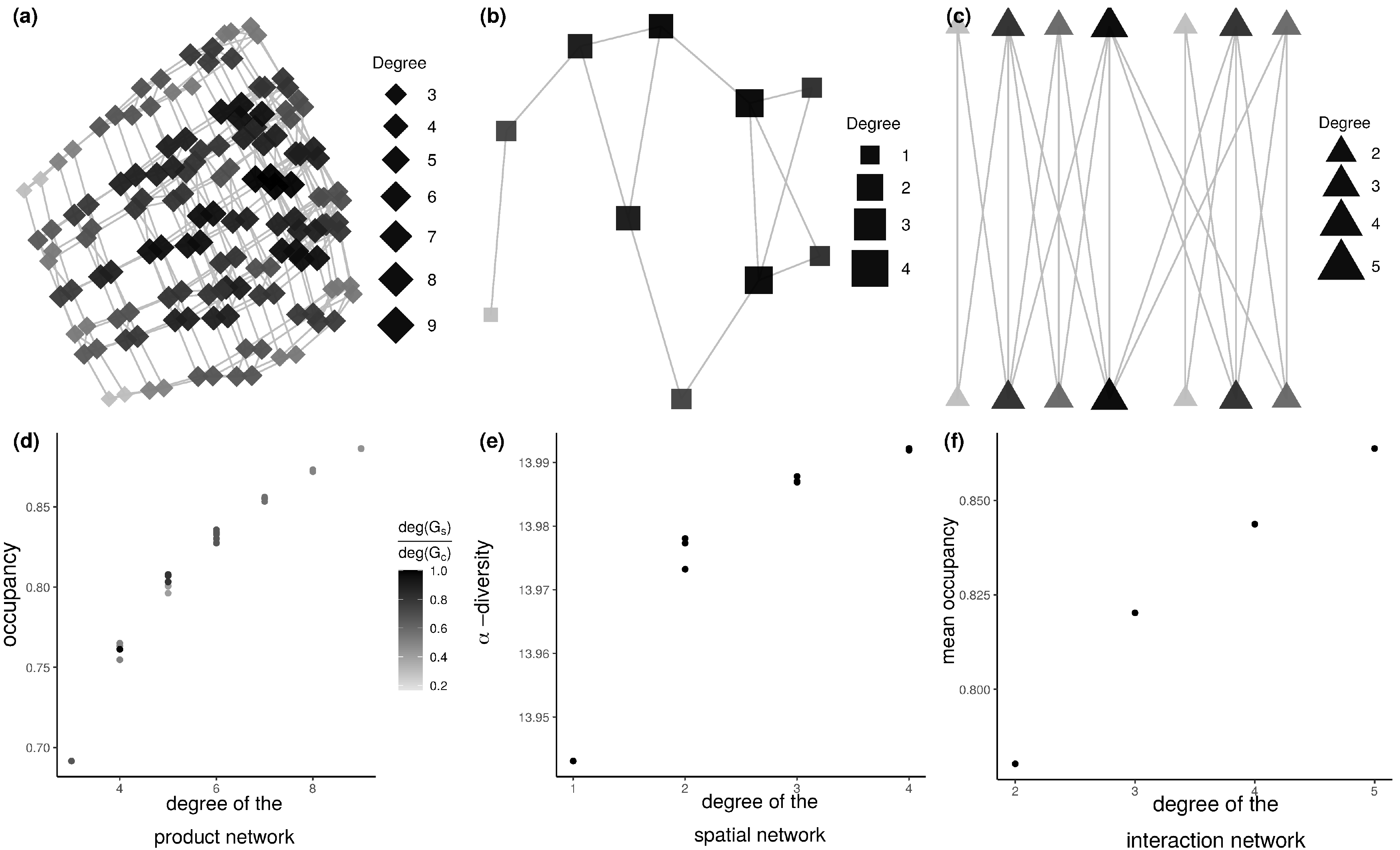}
\caption{
Simulating the dynamics for a given spatial and biotic interaction network with the Levins type submodel.
(a) Colonisation network whose size of the nodes is proportional to their degree and colour indicates the occupancy at equilibrium (grey: low occupancy, black: high occupancy)
(b) Spatial network whose size of the nodes is proportional to their degree and colour indicates the species $\alpha$-diversity at equilibrium (grey: low $\alpha$-diversity, black: high $\alpha$-diversity)
(c)  Mutualistic interaction network whose size of the nodes is proportional to their degree and colour indicates the mean occupancy across the sites at equilibrium (grey: low mean occupancy, black: high mean occupancy)
(d) Relationship between the occupancy at equilibrium and the degree of the node of the product network. Each point of the relationship (corresponding to a node of the product graph) is coloured according to the ratio of the degree of the site in the spatial network over the degree of the focal node in the colonisation network
(e) Relationship between the species $\alpha$-diversity at equilibrium and the degree of the sites in the spatial network
(f) Relationship between the mean occupancy at equilibrium and the degree of the species in the biotic interaction network}
\label{fig:g_c_g_s_g_b_lev}
\end{figure}

 We represented the occupancy of the nodes of the product network (that is the colonisation network in this Levins type submodel) in function of their degree (Fig. \ref{fig:g_c_g_s_g_b_lev}). The occupancy of the nodes of the product network (indexed by a species and a site) increased with the degree of the nodes. Moreover, in this submodel, at a fixed node degree of the product network, the occupancy decreased with the ratio of the degree of the site over the degree of the node of the product network. This means that nodes of the product network that combined a generalist species with a low-connected site have a higher occupancy at equilibrium compared to nodes that combined a specialist species with a highly connected site. We observed the same patterns for the three other submodels (Appendix). From the occupancies at equilibrium, we then computed, species $\alpha$-diversities in each site using the framework developed in \citep{ohlmann2019diversity} with $\eta=2$ (Fig. \ref{fig:g_c_g_s_g_b_lev}). We observed a positive relationship between species $\alpha$-diversity and the degree of the nodes of the spatial network (Fig. \ref{fig:g_c_g_s_g_b_lev}). Mirroring the analysis on the spatial network, we represented the mean occupancy among the sites (Fig. \ref{fig:g_c_g_s_g_b_lev}) and observed a positive relationship between mean occupancy of a species and its degree in the biotic interaction network. 

 \section{Discussion}
 In this paper, we proposed a stochastic, spatially explicit model of mutualistic metacommunities that depends on the structure of spatial and biotic interaction networks, using Dynamic Bayesian Networks. Under a mean-field approximation, we derived a deterministic mutualistic metacommunity model that showed a sharp transition between a state where the metacommunity persisted (i.e. all species have non-null occupancy in all sites), and a state where the metacommunity went extinct (\textit{i.e.} all species had null occupancy in all sites).
 The transition depended on the structure of the interaction and spatial networks and on colonisation and extinction parameters. We defined the metacommunity capacity, a scalar quantity depending on the structure of both networks, as a threshold on colonisation/extinction parameters governing persistence of interacting species, thus extending the single-species concept of metapopulation capacity (\citealt{hanski2000metapopulation}, \citealt{ovaskainen2001spatially}) to a metacommunity context. This threshold has important implications for biodiversity management (\textit{e.g.}, for metapopulations \citealt{groffman2006ecological}), since it helps conservationists to forecast and thus prevent crossing critical thresholds to metacommunity extinction when facing habitat destruction, pollution or other alteration. We extended the framework of metapopulation capacity to the case of a mutualistic metacommunity with a critical extinction threshold that is the same for all species belonging to the metacommunity. Mutualistic interactions thus tangle the fate of the different species involved. This hypothesis is specific to the deterministic model, while local extinctions are possible in our stochastic model. 
 \par Besides conservation, metacommunity capacity can be used to rank different mutualistic systems along a profile of persistence or vulnerability. It can be computed (sum of dominant eigenvalues in the Levins type submodel) from a metanetwork (summarising known potential interactions in the focal region) and from a spatial network built from geographic and environmental space (see \citealt{mimet2013assessing} for further discussions on how to best build spatial networks).
 Importantly, we showed that spatial and interaction networks jointly determine the metacommunity capacity. In other words, any viability statement on a metacommunity (like classic metapopulation viability statements, \textit{e.g.}, \cite{bulman2007minimum}) should be done using both networks, although we should keep in mind that the perceived spatial grain (\textit{i.e.} nodes of the spatial network) and colonisation/extinction parameters might differ among species.
 Interestingly, the dynamics of the node occupancies of the product network allows to follow aggregated quantities like $\alpha$-diversity of the sites or mean occupancy of the species.
 \par
 Our spatially explicit model of mutualistic metacommunities has two notable properties. First, the use of DBNs allows to build models from both the spatial and interaction networks even if their nodes represent distinct entities (site or species) and distinct relationships (geographic proximity or biotic interaction). Second, our mathematical model is built by integrating a metapopulation and a mainland-island interaction model (where species colonise an island influenced by a known meta-network). If metapopulation models are particular cases of the proposed model, the  mainland-island interaction model is not general enough to embed all our proposed models. For example, the trophic theory of island biogeography does not describe dependencies between two consecutive time steps (as in the proposed model) but dependencies, at the current time step, between basal and non-basal species (\citealt{Gravel2011}, \citealt{Massol2017}). It thus much more restrictive compared to what we developed here. 
 Interestingly, while both spatial network and biotic interaction networks are somehow analogous in the DBN formalism, they modulate the colonisation and extinction probabilities in different ways.
If the structure of the networks encode the causal relationships between variables (sites and species), the different parameterisations can encode the way interactions between species and proximity between sites influence the species' probability of presence/absence by using the colonisation and extinction networks.
 We proposed four submodels where the spatial network and the biotic interaction network can affect species colonisation or extinction rates (Table \ref{tab_models}). Other shapes of conditional probabilities could represent other mechanisms, using logic-based rules as suggested in \citep{staniczenko2017linking} and \citep{bohan2011automated}.
 \par
 Our model of mutualistic metacommunity showed a sharp state-transition. Such abrupt transitions are known for community with positive interactions along environmental gradients (\citealt{callaway1997positive}, \citealt{kefi2016can}). We extended these known results for mutualistic metacommunities. Can we expect this for other types of interactions? The assumptions on extinction functions in our model cannot represent non-mutualistic interactions and thus prevent its extension to competitive or multitrophic metacommunities. 
 Regarding competition, competitive exclusion models in communities (\citealt{chesson2000mechanisms}) and metacommunities (\citealt{calcagno2006coexistence}) can lead to several intermediate states between coexistence and extinction of the entire metacommunity. However, competitive interactions along environmental gradients can induced dependencies between species, entailing alternative stable states (\citealt{liautaud2019superorganisms}).
 In the classic Lotka-Volterra deterministic model, conditions on trophic interaction network can lead to states where some of the species goes extinct but not the entire community (\citealt{takeuchi1996global,bunin2017}).
 \cite{wang2021metapopulation} proposed a two species extension of metapopulation capacity with trophic interaction. They consider the metapopulation capacity for the prey and the predator separately. By approximating equilibrium prey occupancy, they compute predator metapopulation capacity. They extend the results to food chain in a hierarchical way. Contrary to the proposed framework, they do not propose a metacommunity capacity but rather a set of metapopulation capacity that depends on each other in hierarchical way. It could be extended towards a trophic metacommunity model in a more general framework in several ways (\citealt{gross2020modern}). However, predicting the outcome of these models from parameters only still poses tough challenges (\citealt{gross2020modern}). In particular, this makes it difficult to establish critical thresholds for conservation science for competitive and trophic metacommunities. Nevertheless, we doubt that a single threshold value governs the fate of many species engaged in several types of interaction with each others as we believe that threshold phenomena occur in multi-interactions metacommunity. Our model should pave the way for a better understanding of properties of spatially realistic trophic and competitive metacommunity models.
 
\par
\section*{Acknowledgements}
We thank Fabien Laroche for insightful comments and references on metapopulation and metacommunity models. This research was funded by the French Agence Nationale de la Recherche (ANR) through the GlobNet (ANR-16-CE02-0009) and EcoNet (ANR-18-CE02-0010) projects and from ‘Investissement d'Avenir’ grants managed by the ANR (Montane: ANR‐10‐LAB‐56).  

\appendix
\section{Appendix: proofs and details on the model}
\subsection{The $nm$-intertwined model}
The approximation is derived from \cite{bianconi2018multilayer} and \cite{van2011n}. The aim is to study the dynamics of occupancy of each species $j$ in each site $i$: $p_{ij}(t)=\EX(X_{ij}^t)$). For all $i$ and $j$, we have
\begin{equation}
        p_{i,j}(t+1)=\EX((1-X_{i,j}^t)(\epsilon + (1-\epsilon)(1-(1-c)^{\sum_{(k,l) \in N_{G_c}(i,j)} X^t_{k,l}})))+\EX(X_{i,j}^t (1-e(1 - \frac{\sum_{(k,l) \in N_{G_e}(i,j)} X^t_{k,l}}{1+deg_{G_e}((i,j))})))
\end{equation}

 This approach leads to a hierarchy of equations that cannot be solved (i.e. we need to consider $\EX(X^t_{1,1},...,X^t_{m,n})$ to find a solution to the system). A drastic approximation consists in the mean field approximation, for any sequence of indices $n(1),n'(1);...,n(r),n'(r')$, we assume :
\begin{equation}
     \EX(X^t_{n(1),n'(1)},...,X^t_{n(r),n'(r')}) \simeq \EX(X^t_{n(1),n'(1)})...\EX(X^t_{n(r),n'(r')})  
\end{equation}
\begin{equation}
    p_{i,j}(t+1)=(1-p_{i,j}(t))(\epsilon + (1-\epsilon)(1-(1-c)^{\sum_{(k,l) \in N_{G_c}(i,j)} p_{kl}(t)})+(1-e(1 - \frac{\sum_{(k,l) \in N_{G_e}(i,j)} p_{k,l}(t)}{1+deg_{G_e}((i,j))}))p_{i,j}(t)
\end{equation}
We assume that $c=o(1)$, $e=o(1)$ and $\epsilon=o(c)$, a Taylor expansion at order 1 with set $M_{i,j}:=1+deg_{G_e}((i,j))$ leads to:
\begin{equation}
    p_{i,j}(t+1)=(1-p_{i,j}(t))(\epsilon + (1-\epsilon)(c\sum_{(k,l) \in N_{G_c}(i,j)} p_{k,l}(t))+(1-e+e\sum_{(k,l) \in N_{G_e}(i,j)} p_{k,l}(t))/M_{i,j})p_{i,j}(t)
\end{equation}
\begin{equation}
    p_{i,j}(t+1)=(1-p_{i,j}(t))(c\sum_{(k,l) \in N_{G_c}(i,j)} p_{k,l}(t))+(1-e+e\sum_{(k,l) \in N_{G_e}(i,j)} p_{k,l}(t))/M_{i,j})p_{i,j}(t)
\end{equation}
We introduce a single index $v$ for the nodes of the product networks and get:
\begin{equation}
    p_{v}(t+1)=(1-p_{v}(t))(c\sum_{u \in N_{G_c}(v)} p_{u}(t))+(1-e+e\sum_{u \in N_{G_e}(v)} p_{u}(t))/M_u)p_{v}(t)
\end{equation}
\begin{equation}
    p_{v}(t+1)-p_{v}(t)=(1-p_{v}(t))(c\sum_{u \in N_{G_c}(v)} p_{u}(t))-e(1-\sum_{u \in N_{G_e}(v)} p_{u}(t))/M_u)p_{v}(t)
\end{equation}

\begin{equation}
    p_v(t+1)-p_v(t)=C_v(\mathbf{p(t)})(1-p_v(t)))-E_v(\mathbf{p(t)})(p_v(t))
    \label{eq:master_eq}
\end{equation}
where $C_v(\mathbf{p(t)})=c\sum_{u}[A_c]_{v,u}p_{u}(t)$ and $E_v(\mathbf{p(t)})=e(1-\sum_{u}[A_e]_{v,u}p_{u}(t)/M_u)$. \\
\begin{equation}
    \mathbf{p(t+1)}-\mathbf{p(t)} = c(\mathbf{A_c}\mathbf{p(t)})\odot (\mathbf{1} - \mathbf{p(t)}) -e(\mathbf{1} - (\mathbf{D_e} + \mathbf{I}_{nm})^{-1}\mathbf{A_e}\mathbf{p(t)}) \odot \mathbf{p(t)}
\end{equation}

where $\odot$ denotes the element-wise product, $D_e$ denotes the indegree matrix of $G_e$ and $\mathbf{I}_{nm}$ the identity matrix of dimension $nm$.

\subsection{Proof of proposition~\ref{prop:irreducible}}
We need to show that the four submodels are irreducible. \\
We have:
\begin{equation}
    g_v(\mathbf{p}) = \frac{\sum_{u}[A_{c}]_{v,u}p_{u}}{1-\sum_{u}[A_{e}]_{v,u}p_{u}/M_u}
\end{equation}
\begin{equation}
    \frac{\partial g_v}{\partial p_u}(\mathbf{p}) = \frac{[A_{c}]_{v,u}(1+\sum_{k}\frac{[A_{e}]_{v,k}p_k}{M_k})-[A_{e}]_{v,u}\sum_{k}\frac{[A_{c}]_{v,k}p_k}{M_k}}{(1-\sum_{u}[A_{e}]_{v,u}p_{u}/M_u)^2}
\end{equation}
\\
For the Levins type model:
\begin{equation}
     \frac{\partial g_v}{\partial p_u}(\mathbf{0}) = [A_{c}]_{v,u}
\end{equation}
And since, $G_s$ and $G_b$ are both strongly connected and $G_c=G_s \square G_b$, $G_c$ is strongly connected and $\mathbf{J}$ (built from the jacobian matrix of $g$) is irreducible.\\ \\
For the separated effect model, $E(G_e) \cap E(G_c)=\emptyset$. So $\sum_{v,u}[A^e]_{v,u}[A^c]_{v,u}=0$. Then,
 \begin{itemize}
        \item if $[A_{c}]_{v,u}=0$, then 
        \begin{equation}
        \frac{\partial g_v}{\partial p_u}(\mathbf{p})=-\frac{[A_{e}]_{v,u}{\sum_{k}[A_{c}]_{v,k}p_k/M_k}}{E_v(\mathbf{p})^2}
        \end{equation} 
        since $ \forall (v,k), [A_{c}]_{v,k} \geq 0$ and $[A_{e}]_{v,k} \geq 0$, then:\\
        \begin{equation}
        g_{v,u}(\mathbf{p}) > 0 \Leftrightarrow (v,u) \in E(G_e)
        \end{equation}
        \item if $[A_{e}]_{v,u}=0$, then:
        \begin{equation}
        g_{v,u}(\mathbf{p}) > 0 \Leftrightarrow (v,u) \in E(G_c)
        \end{equation}
    \end{itemize}
So,\\
\begin{equation}
g_{v,u}(\mathbf{p}) > 0 \Leftrightarrow (v,u) \in E(G_c) \cup E(G_e)
\end{equation}
and since $E(G_c) \cup E(G_e)$ is the edge set of $G_s \square G_b$, that is strongly connected, it follows that $\mathbf{J}$ is irreducible in this case.
\\ 
Similarly, the combined effect model and the rescue effect model are irreducible.\\ \\

\subsection{Computation of $\lambda_I$}
As provided in the main text, for the Levins type submodel, $\lambda_I = \lambda_M = \Lambda_s + \Lambda_b$.
We now compute the $\lambda_I$ for the three other submodels. \\
We first compute the Jacobian matrix of $\mathbf{p} \mapsto
g(\mathbf{p})$ evaluated in $\mathbf{p}=0$.
We have 
\begin{equation}
    \frac{\partial g_v}{\partial p_u}(\mathbf{0}) =[A_{c}]_{v,u}
\end{equation}
$\lambda_I$ is the dominant eigenvalue of $\left( \frac{\partial g_v}{\partial p_u}(\mathbf{0}) \right)_{u,v}$
\begin{itemize}
    \item Separated effect submodel\\
    For this submodel, $A_{c}=A_s$, it follows $\lambda_I=\Lambda_s$
    \item Combined effect submodel\\
    For this submodel, $A_{c}=A_s \otimes I_m + I_n \otimes A_b$, it follows $\lambda_I=\Lambda_s$
    \item Rescue effect submodel\\
    For this submodel, $A_{c}=A_s$, it follows $\lambda_I=\Lambda_s$
\end{itemize}
\subsection{Computation of $\lambda_M$}
\begin{figure}[h!]
\centering
\includegraphics[width=13cm]{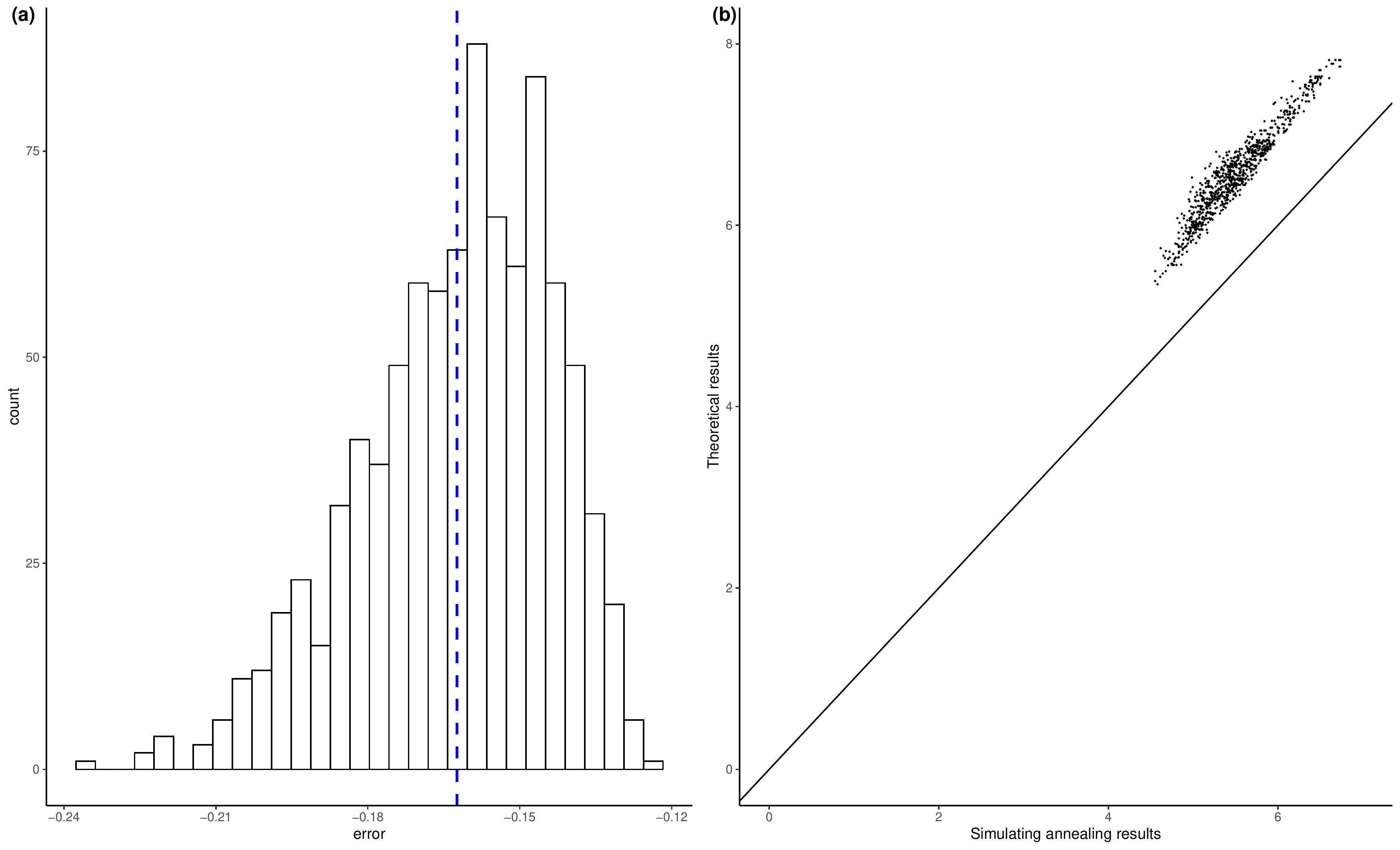}
\caption{(a) Distribution of the relative error in the estimation of the metacommunity capacity (b) Relation between the metacommunity persistence capacity computing using a simulating annealing algorithm and the theoretical metacommunity capactity for the Levins type submodel}
\label{fig:SA_error_lev}
\end{figure}

In order to compute $\lambda_M$ for the combined effect submodel, the separated effect model and the rescue effect submodel where the components of $\mathbf{g}$ are not concave, we used a simulated annealing algorithm. We used the result of the iterative procedure described in Appendix D of \citealt{ovaskainen2001spatially} as starting point.\\
The code to compute the metacommunity capacity in the different models is available at: \url{https://gitlab.com/marcohlmann/metacommunity_theory}.\\
We assessed the performance of the method on the Levins type model on the simulated data, since we know analytically the metacommunity capacity in this case.
We used 20000 time steps on the $900$ different networks for the $4$ submodels. The maximum is not reached (Fig. S\ref{fig:SA_error_lev}a) but the there is a strong correlation (0.955) between the estimated metacommunity capacities and the theoretical metacommunity capacities (Fig. S\ref{fig:SA_error_lev}b), allowing so comparison of the metacommunity capacities among the different network structures.

\section{Appendix: detail on the simulation}
\subsection{Spatial networks}
In order to mimic fragmentation of the landscape, we sampled spatial networks (10 nodes) using Erd\H{o}s-Renyi model and a block model.
For the Erd\H{o}s-Renyi model, the probability of connection was $C=0.25$ and we kept connected networks only.
For the block model, we partitioned in two groups of equal sizes, $p$ and $q$, with a matrix of probability of connection, $\mathbf{\Pi}$, given by:
$$
\begin{blockarray}{ccc}
p & q \\
\begin{block}{(cc)c}
  \frac{7C}{4} &  \frac{C}{4} & p \\
   \frac{C}{4} &  \frac{7C}{4} & q \\
\end{block}
\end{blockarray}
$$
where $C=0.25$.\\
The overall probability of connection in the network is :
\begin{equation}
        \mathbb{P}(i \leftrightarrow j) = \sum_{k\in\{p,q\},l\in\{p,q\}}Pr(i \leftrightarrow j | i \in k, j \in l)Pr(i \in k)Pr(j \in l)
\end{equation}

\begin{equation}
        \mathbb{P}(i \leftrightarrow j) = \frac{1}{4}(\frac{7C}{4}+\frac{C}{4}+\frac{C}{4}+ \frac{7C}{4})
\end{equation}

\begin{equation}
 \mathbb{P}(i \leftrightarrow j) = C   
\end{equation}

\begin{figure}[h!!!]
\centering
\includegraphics[width=10cm]{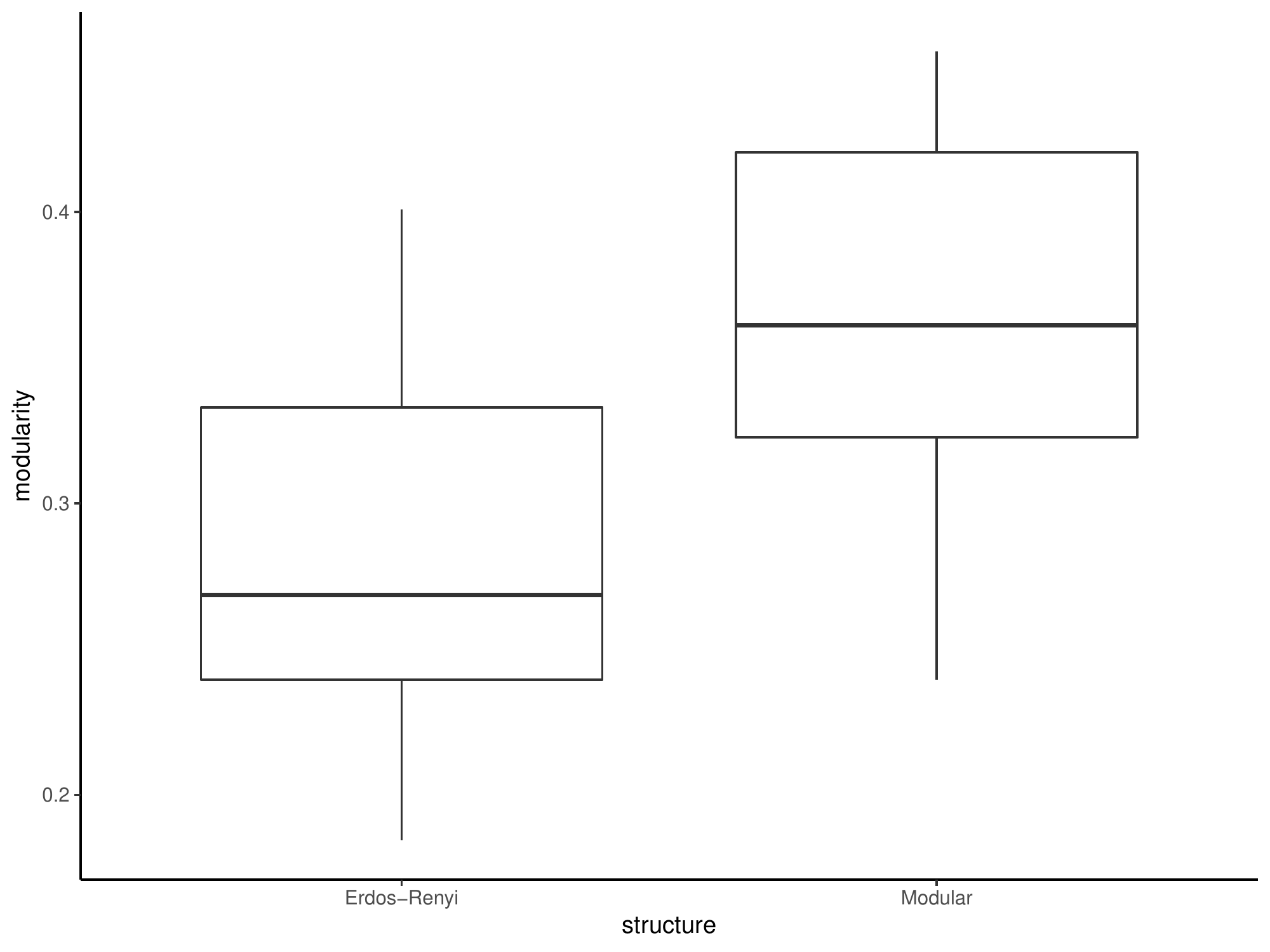}
\caption{Distribution of the modularity of the spatial networks over the $15$ replicates for the Erd\H{o}s-Renyi strucutre and the modular structure}
\label{fig:modularity}
\end{figure}

So the expected value of connectance for all spatial networks is the same despite different modularity values (Fig. S\ref{fig:modularity}).

\subsection{Biotic interaction networks}
We first generated random undirected network with various shapes of the degree distribution using the function \textit{sample\_fitness\_pl} implemented in the R package \textit{igraph} \citep{igraph}. We generated Erd\H{o}s-Renyi networks and networks with a degree distribution given by a power-law.\\
We only kept connected networks. On the random network $G$ sampled ($\mathbf{A}$ is its adjacency matrix), we build a bipartite network $G_{bip}$ with adjacenncy matrix $\mathbf{A_{bip}}$ as:
\begin{equation}
     \mathbf{A_{bip}} = \mathbf{A_2} \otimes (\mathbf{A} + \mathbf{I_n})
\end{equation}

where $\mathbf{A_2}$ is the adjacency matrix of an undirected graph made of two nodes and a single edge between these two nodes. By doing so, all the sampled undirected bipartite networks are strongly connected.
\subsection{Results}
\begin{figure}[h!!!]
\centering
\includegraphics[width=10cm]{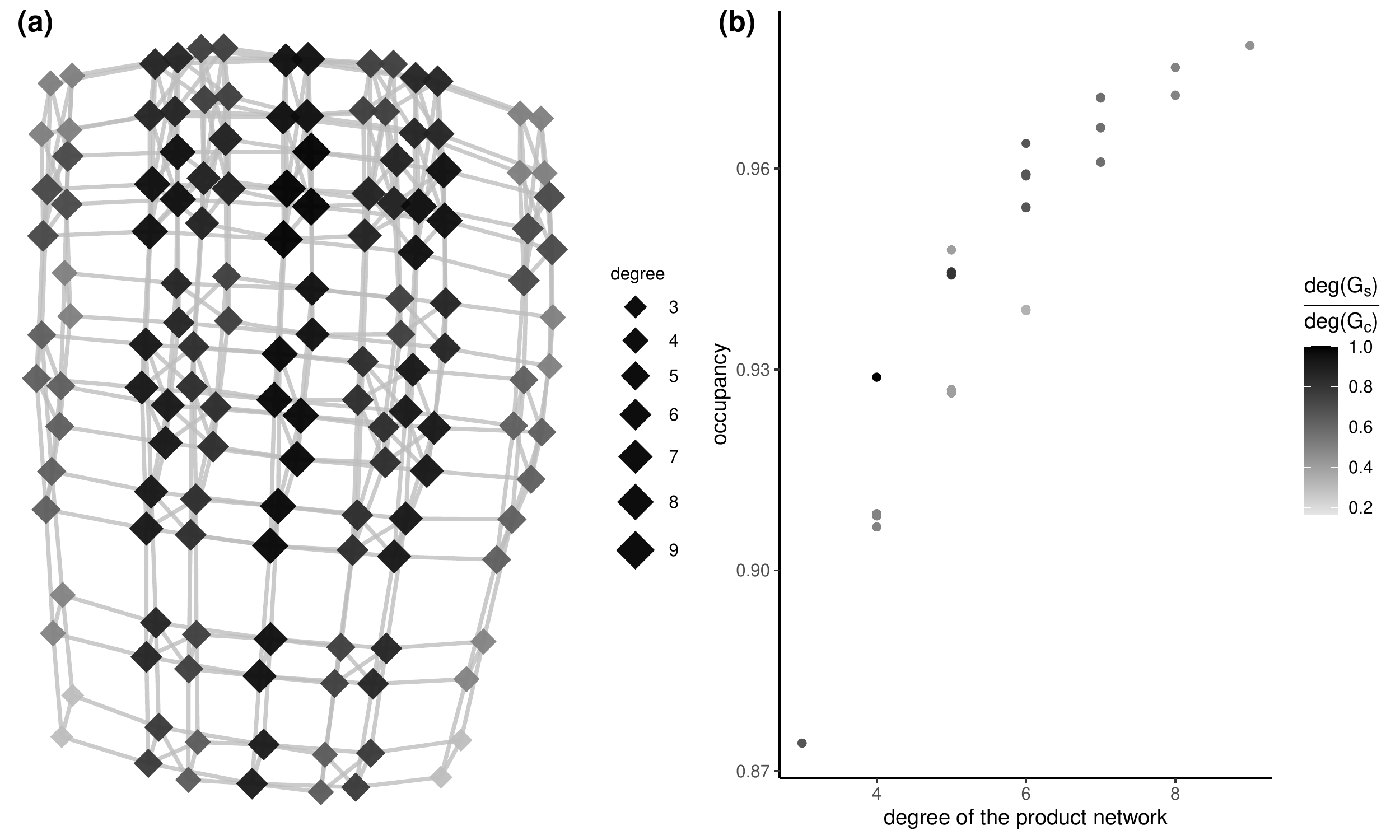}
\caption{Simulating the dynamics for a given spatial and biotic interaction network with the combined effect submodel. (a) colonisation network whose size of the nodes is proportional to their degree and colour indicates the occupancy at equilibirum (grey: low occupancy, black: high occupancy). (b) Relationship between the occupancy at equilibrium and the degree of the node of the product network. Each point of the relationship (corresponding to a node of the product graph) is coloured according to the ratio of the degree of the site in the spatial network over the degree of the focal node in the colonisation network.}
\label{fig:g_c_lev}
\end{figure}

\begin{figure}[h!!!]
\centering
\includegraphics[width=10cm]{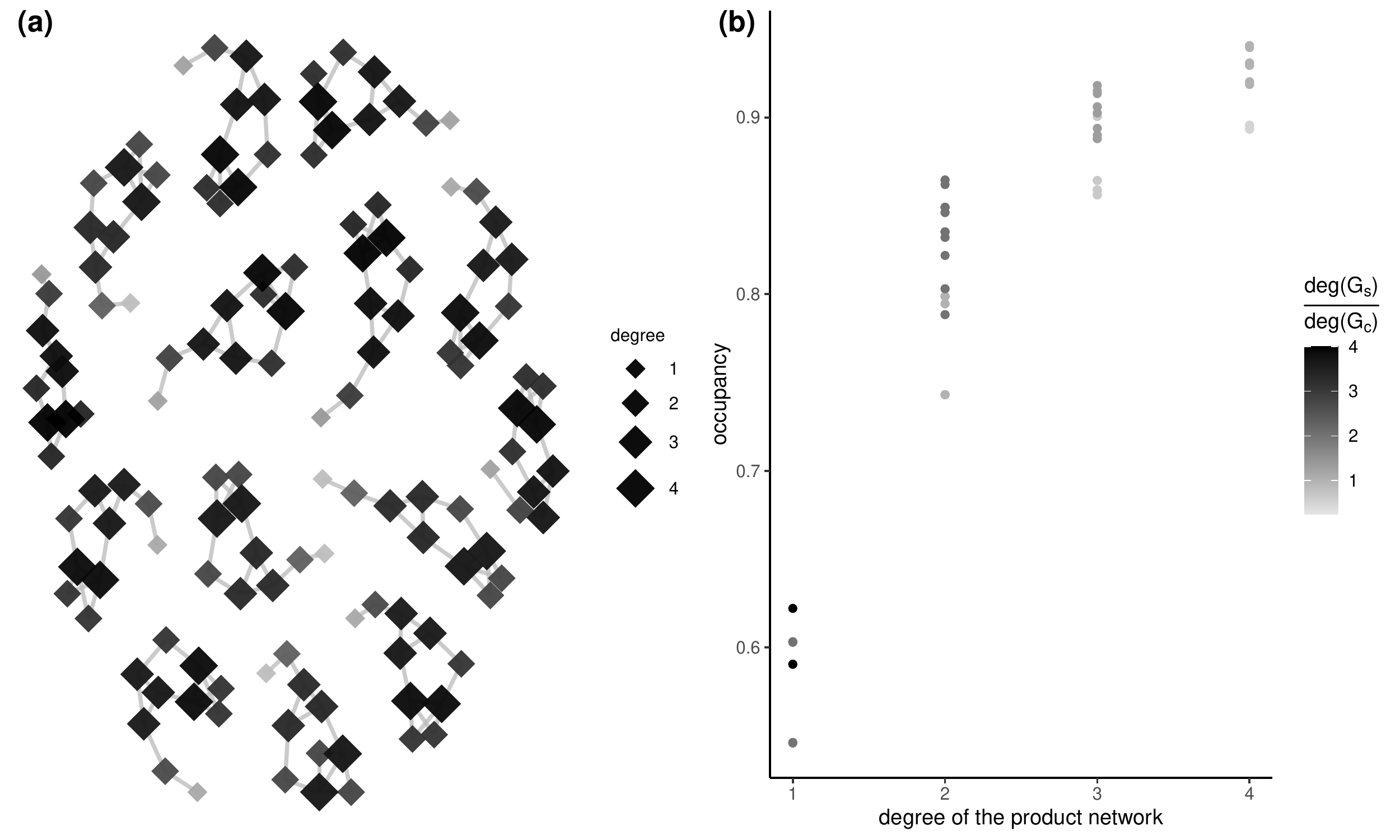}
\caption{Simulating the dynamics for a given spatial and biotic interaction network with the separated effect submodel. (a) colonisation network whose size of the nodes is proportional to their degree and colour indicates the occupancy at equilibirum (grey: low occupancy, black: high occupancy). (b) Relationship between the occupancy at equilibrium and the degree of the node of the product network. Each point of the relationship (corresponding to a node of the product graph) is coloured according to the ratio of the degree of the site in the spatial network over the degree of the focal node in the colonisation network.}
\label{fig:g_c_sep}
\end{figure}

\begin{figure}[h!!!]
\centering
\includegraphics[width=10cm]{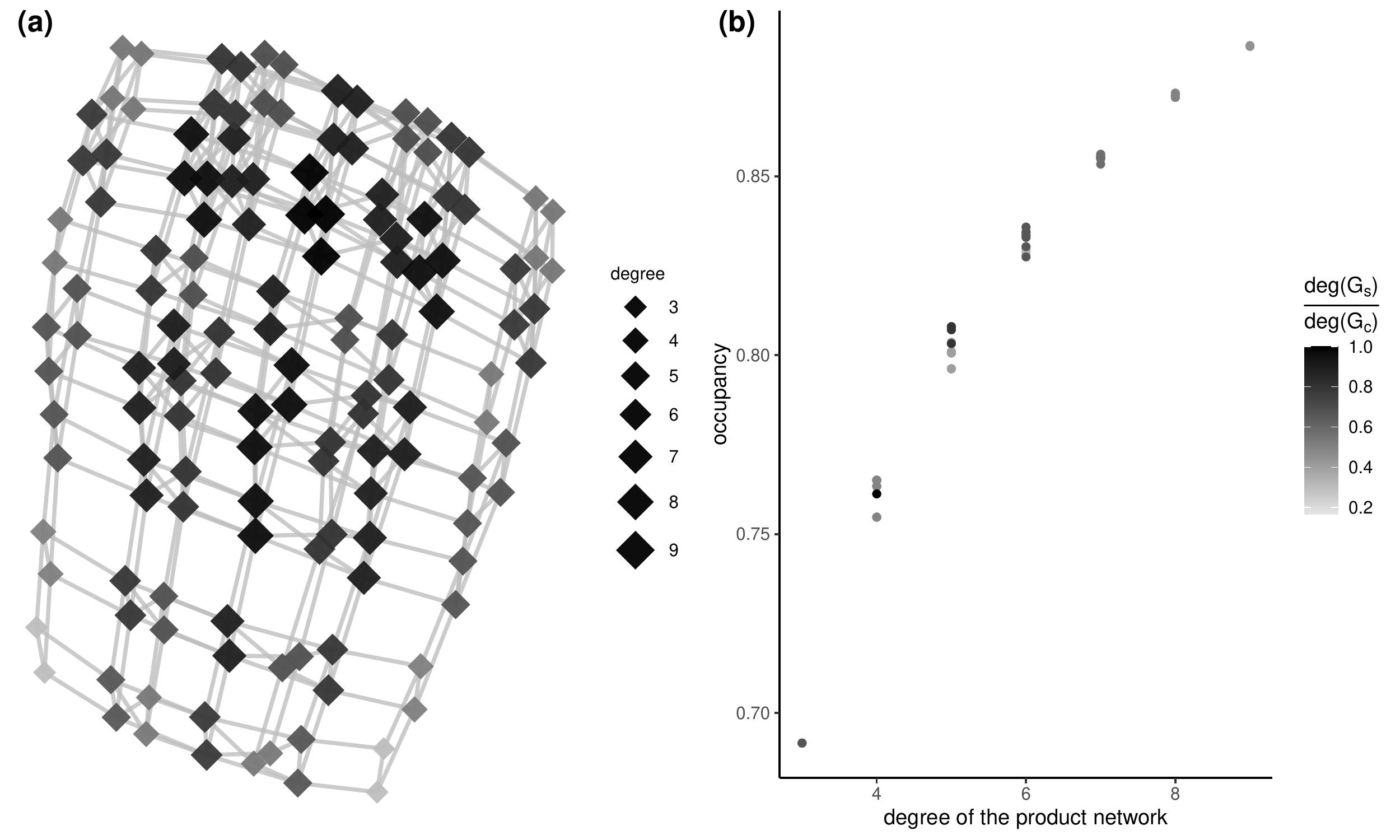}
\caption{Simulating the dynamics for a given spatial and biotic interaction network with the rescue effect submodel. (a) colonisation network whose size of the nodes is proportional to their degree and colour indicates the occupancy at equilibirum (grey: low occupancy, black: high occupancy). (b) Relationship between the occupancy at equilibrium and the degree of the node of the product network. Each point of the relationship (corresponding to a node of the product graph) is coloured according to the ratio of the degree of the site in the spatial network over the degree of the focal node in the colonisation network.}
\label{fig:g_c_resc}
\end{figure}

\begin{figure}[h!!!]
\centering
\includegraphics[width=10cm]{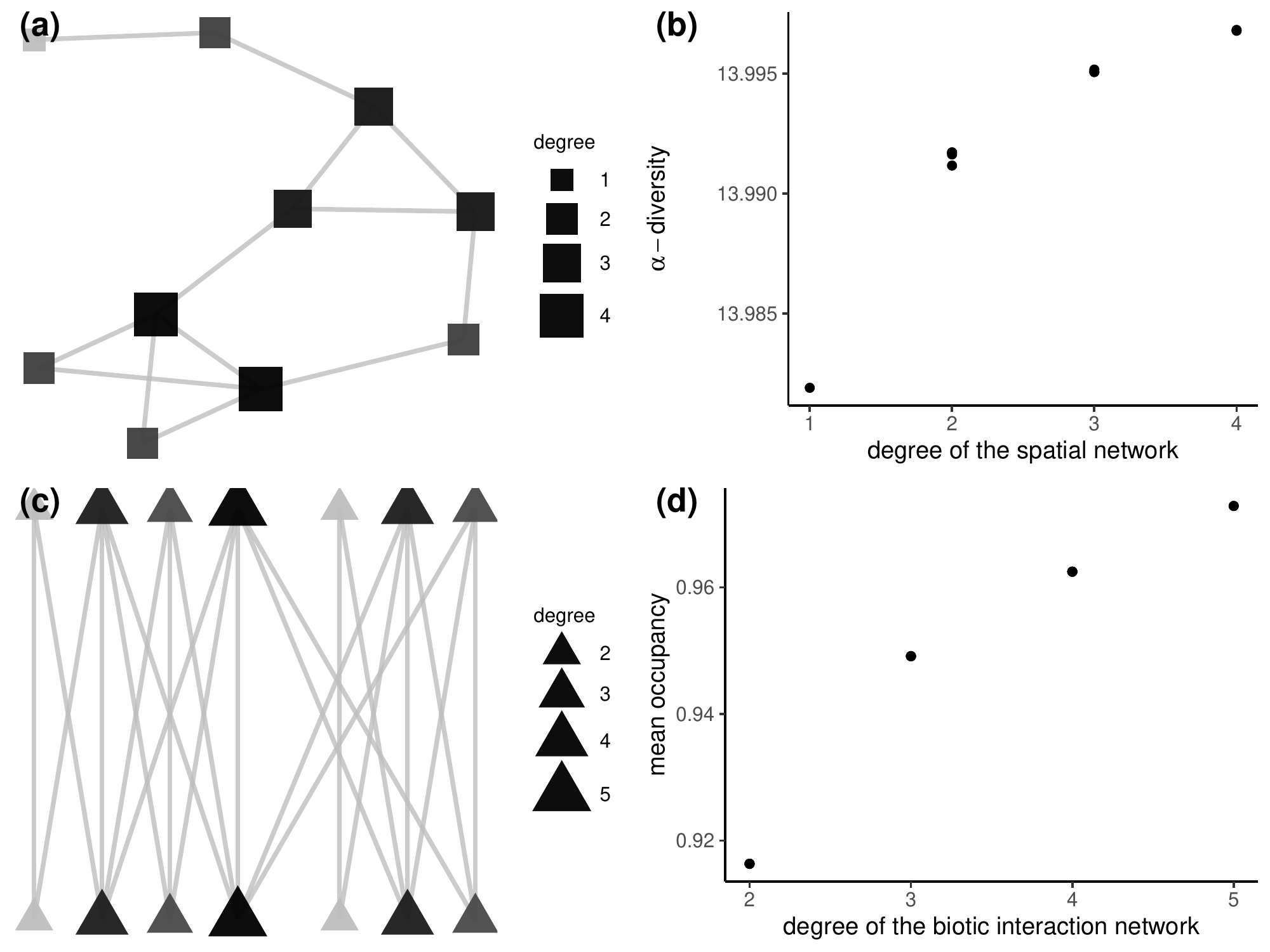}
\caption{Aggregated statistics from occupancy at equilibrium for the combined effect submodel in the spatial network and the biotic interaction network. (a) Spatial network whose size of the nodes is proportional to their degree and colour indicates the $\alpha$-diversity at equilibrium (grey: low $\alpha$-diversity, black: high $\alpha$-diversity). (b) Relationship between the $\alpha$-diversity at equilibrium and the degree of the sites in the spatial network. (c)  Biotic interaction network whose size of the nodes is proportional to their degree and colour indicates the mean occupancy across the sites at equilibrium (grey: low $\alpha$-diversity, black: high $\alpha$-diversity). (d) Relationship between the mean occupancy at equilibirum and the degree of the species in the biotic interaction network.}
\label{fig:g_s_g_b_lev}
\end{figure}

\begin{figure}[h!!!]
\centering
\includegraphics[width=10cm]{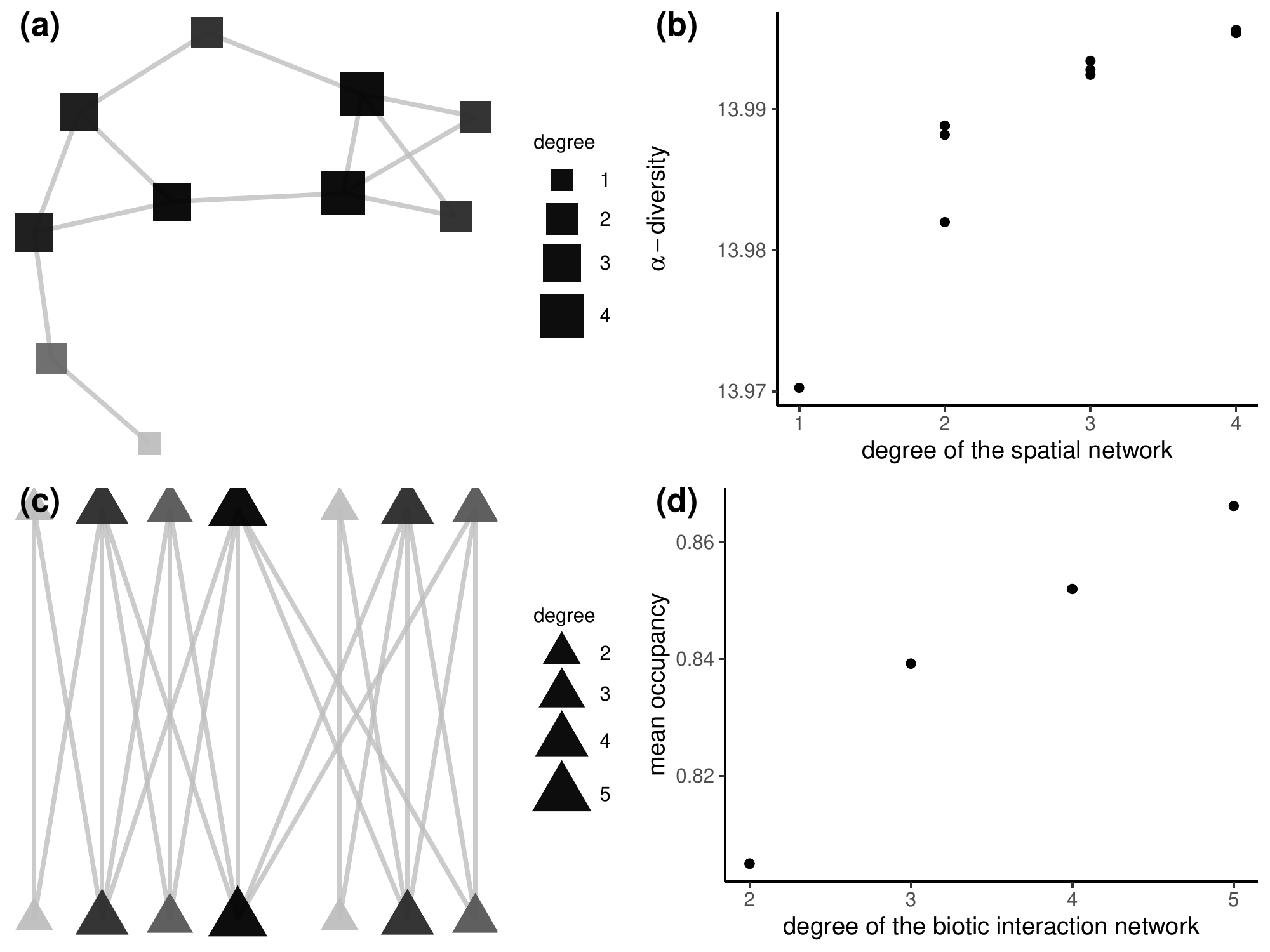}
\caption{Aggregated statistics from occupancy at equilibrium for the separated effect submodel in the spatial network and the biotic interaction network. (a) Spatial network whose size of the nodes is proportional to their degree and colour indicates the $\alpha$-diversity at equilibrium (grey: low $\alpha$-diversity, black: high $\alpha$-diversity). (b) Relationship between the $\alpha$-diversity at equilibrium and the degree of the sites in the spatial network. (c)  Biotic interaction network whose size of the nodes is proportional to their degree and colour indicates the mean occupancy across the sites at equilibrium (grey: low $\alpha$-diversity, black: high $\alpha$-diversity). (d) Relationship between the mean occupancy at equilibirum and the degree of the species in the biotic interaction network.}
\label{fig:g_s_g_b_sep}
\end{figure}

\begin{figure}[h!!!]
\centering
\includegraphics[width=10cm]{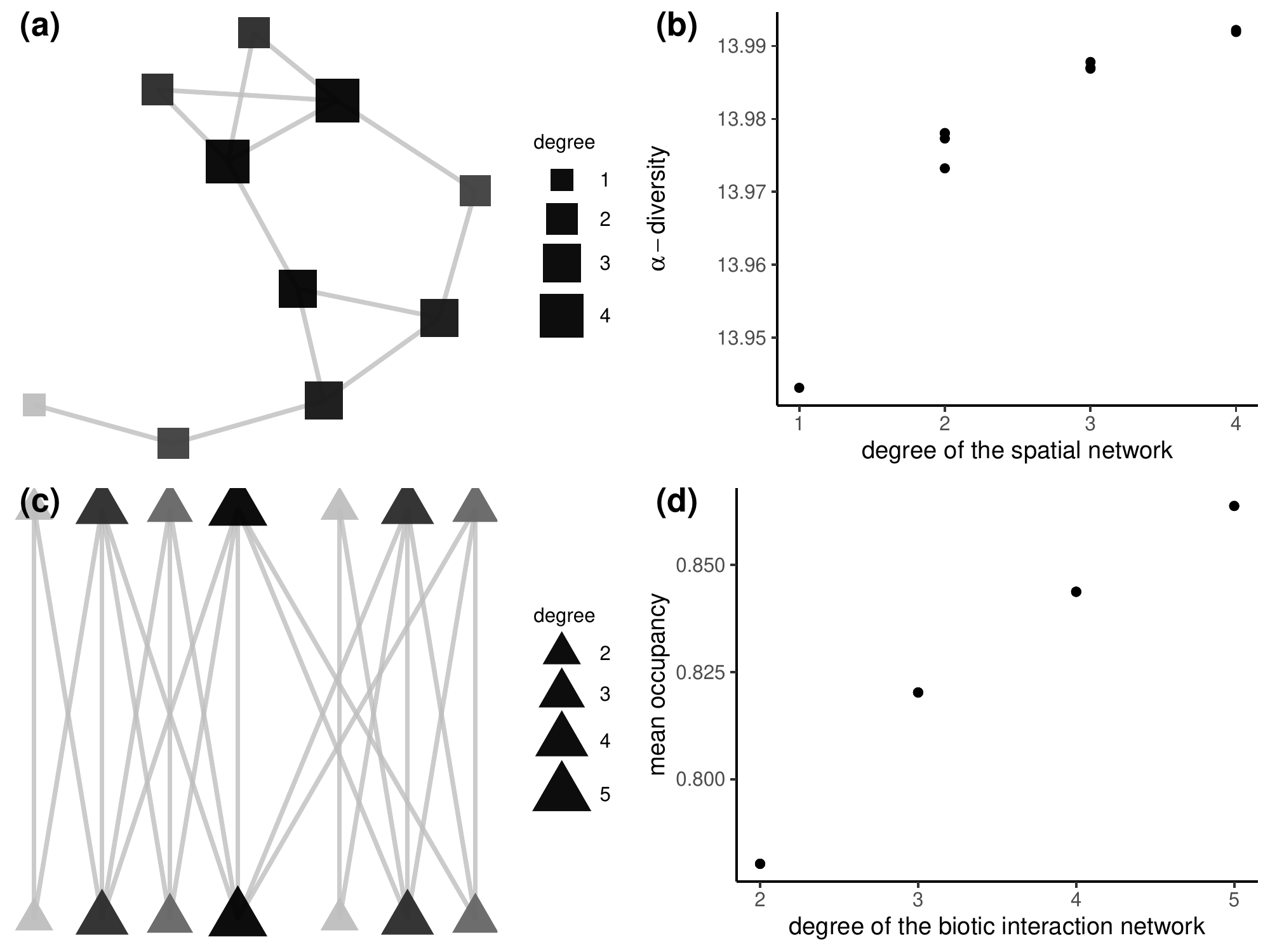}
\caption{Aggregated statistics from occupancy at equilibrium for the rescue effect submodel in the spatial network and the biotic interaction network. (a) Spatial network whose size of the nodes is proportional to their degree and colour indicates the $\alpha$-diversity at equilibrium (grey: low $\alpha$-diversity, purple: black $\alpha$-diversity). (b) Relationship between the $\alpha$-diversity at equilibrium and the degree of the sites in the spatial network. (c)  Biotic interaction network whose size of the nodes is proportional to their degree and colour indicates the mean occupancy across the sites at equilibrium (grey: low $\alpha$-diversity, black: high $\alpha$-diversity). (d) Relationship between the mean occupancy at equilibirum and the degree of the species in the biotic interaction network.}
\label{fig:g_s_g_b_resc}
\end{figure}

We simulated the dynamic (as presented in the main text for the Levins type submodel) for the combined effect submodel (Fig. S\ref{fig:g_c_lev}, Fig. S\ref{fig:g_s_g_b_lev}), for the seperated effect submodel (Fig. S\ref{fig:g_c_sep}, Fig. S\ref{fig:g_s_g_b_sep}) and for the rescue effect submodel (Fig. S\ref{fig:g_c_resc}, Fig. S\ref{fig:g_s_g_b_resc}).

\subsection{Robustness of metacommunity capacity estimation}
\begin{figure}[h!!!]
\centering
\includegraphics[width=15cm]{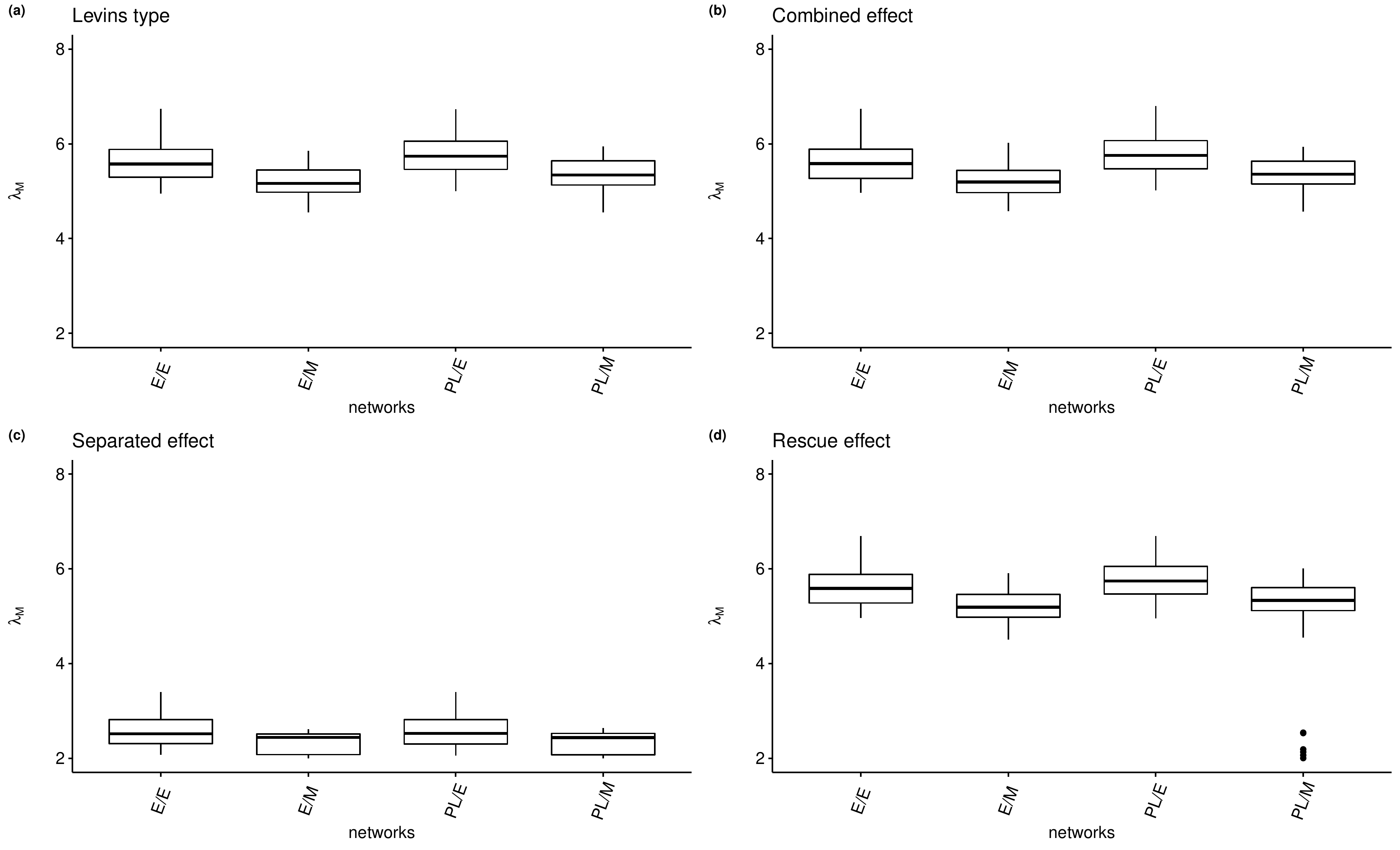}
\caption{Boxplot representing distributions of $\lambda_M$ for each combination of structure and each submodel. E: Erd\H{o}s-Renyi, PL: Power-Law, M: Modular}
\label{fig:boxplot}
\end{figure}

\begin{figure}[h!!!]
\centering
\includegraphics[width=15cm]{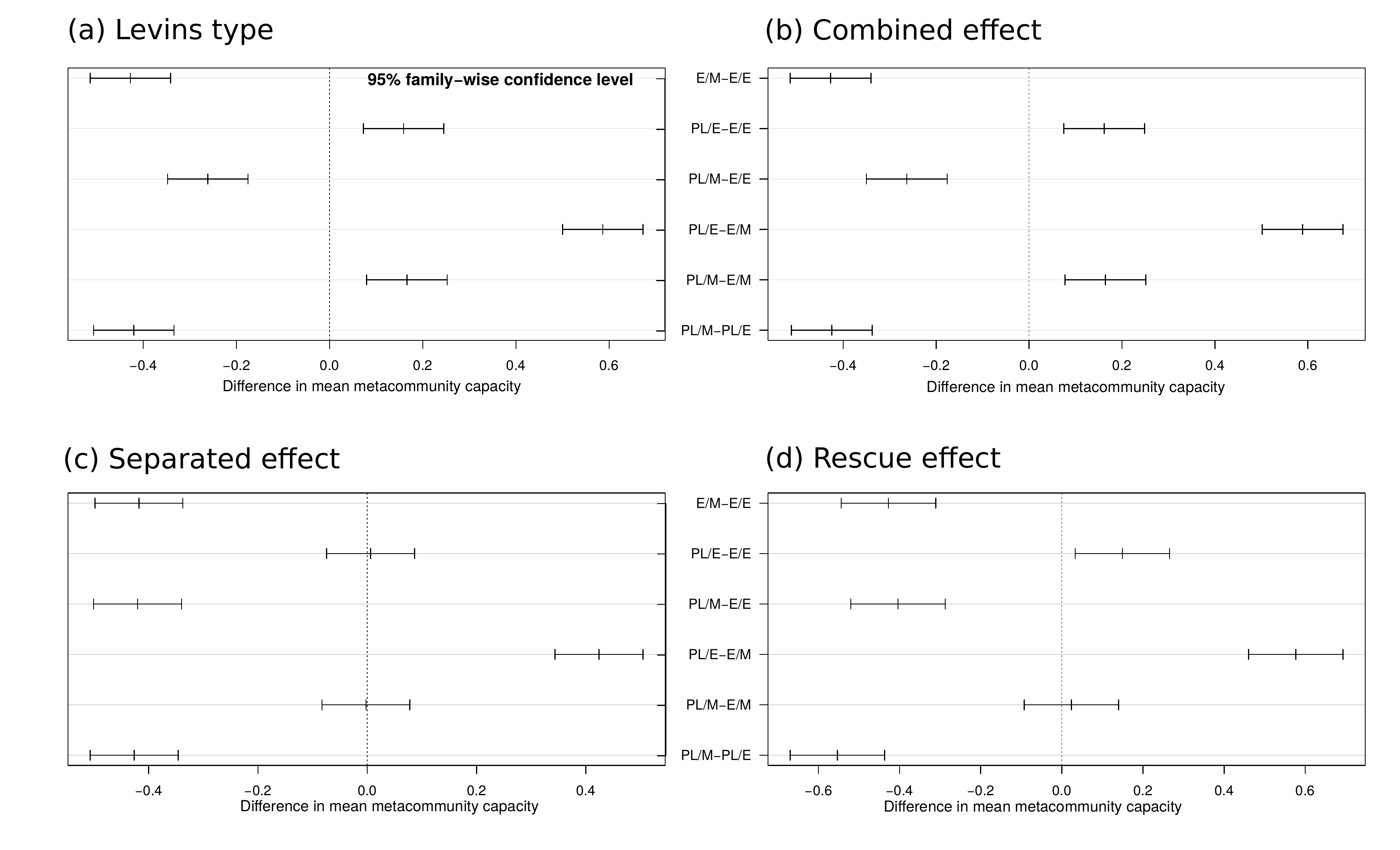}
\caption{Tukey plot representing the confidence intervals of the difference in mean metacommunity capacity per pairs of structures. E: Erd\H{o}s-Renyi, PL: Power-Law, M: Modular}
\label{fig:tukey}
\end{figure}

We analysed the robustness of the estimation of $\lambda_M$ for the four different structures for each submodel. We described the distribution of $\lambda_M$ ($225$ samples per combination of structure for each model) using a boxplot (Fig. S\ref{fig:boxplot}). Morever, we used a Tukey test to estimate the confidence intervals of the difference in mean metacommunity capacity per pairs of structures (Fig. S\ref{fig:tukey}). For the Levins type and combined effect model, all differences in mean $\lambda_M$ were statistically different of $0$. For the seperated effect and rescue effect model, difference in mean $\lambda_M$ of PL/E-E/E (PL: Power-Law, E: Erd\H{o}s-Renyi, M: Modular) and PL/M-E/M were statistically not different from $0$. It means that, for these two models, whatever the structure of the spatial network (Modular or Erd\H{o}s-Renyi), mean $\lambda_M$ was comparable for a power-law or Erd\H{o}s-Renyi biotic interaction network.

 \newpage
 
\bibliographystyle{format}
\bibliography{bib_p.bib}

\end{document}